\documentclass[10pt, aps,pra,amsmath,superscriptaddress,twocolumn,longbibliography,nofootinbib]{revtex4-2}

\usepackage{graphicx}
\usepackage{amsmath}
\usepackage{amsfonts}
\usepackage{amssymb}
\usepackage{mathrsfs}
\usepackage{braket}
\usepackage{bm}
\usepackage{multirow}
\usepackage{color}
\usepackage[normalem]{ulem}
\usepackage{mathrsfs}
\usepackage{mathtools}
\usepackage{float}
\usepackage{enumitem}
\usepackage{wasysym}

\usepackage{dsfont}
\usepackage{graphicx}
\usepackage{xcolor}

\makeatletter

\newcommand{\Rmnum}[1]{\expandafter\@slowromancap\romannumeral #1@}
\makeatother

\def\ket#1{|\, #1\,\rangle}
\def\bra#1{\langle\,#1\,|}

\begin{document}

\title{Cooper-pair splitters as circuit elements for realizing
  topological superconductors}

\author{Guilherme Delfino}
\affiliation{Department of Physics, Boston University, Boston, Massachusetts 02215, USA}

\author{Dmitry Green}
\affiliation{Department of Physics, Boston University, Boston, Massachusetts 02215, USA}
\affiliation{AppliedTQC.com, ResearchPULSE LLC, New York, NY 10065, USA}

\author{Saulius Vaitiek\.enas}
\affiliation{Center for Quantum Devices, Niels Bohr Institute,
University of Copenhagen, 2100 Copenhagen, Denmark}

\author{Charles M. Marcus}
\affiliation{Center for Quantum Devices, Niels Bohr Institute,
University of Copenhagen, 2100 Copenhagen, Denmark}
\affiliation{Department of Physics, University of Washington, Seattle, Washington 98195, USA}
\affiliation{Materials Science and Engineering, University of Washington, Seattle, Washington 98195, USA}

\author{Claudio Chamon}
\affiliation{Department of Physics, Boston University, Boston, Massachusetts 02215, USA}

\date{\today}

\begin{abstract}
  Advances in materials and fabrication of superconducting devices allows the exploration of novel quantum effects in synthetic superconducting systems beyond conventional Josephson junction arrays. As an example, we introduce a new circuit element, the Y-splitter, a superconducting loop with three leads and three Josephson junctions, smaller or comparable in size to the superconducting coherence length of the material. By tuning magnetic flux through an array of Y-splitters, Cooper-pair transport can be made to interfere destructively, while spatially separated split Cooper pairs  propagate coherently. We consider an array of Y-splitters connected in a two-dimensional star [Archimedean (3,$12^2$)] geometry, deformable into the kagome lattice, and find a rich phase diagram that includes topological superconducting phases with Chern numbers $\pm 2$. Experimental realization  appears feasible.
\end{abstract}

\maketitle


{\it Introduction:} The suppression of dissipation in superconductors is key to building electrical circuits that behave quantum mechanically, making these circuits leading contenders for quantum computing architectures and topological phases~\cite{Devoret2004,Wendin2017, Kjaergaard2020, Blais2021,Lesser2024, riwar2016multi}. To date, the collection of non-dissipative superconducting circuit elements includes capacitors, inductors, and Josephson junctions. In describing these circuits theoretically, the underlying pairing of electrons is concealed; aside from occasional factors of two in formulas, electron pairing is assumed then forgotten, except insofar as the separation of pairs is regarded as an undesirable source of decoherence and parity non-conservation.   

In parallel to the advances in superconducting circuits, there has been considerable development within mesoscopic physics of superconductors. Theoretical and experimental effort has focused on the controlled splitting of Cooper pairs in a variety of platforms, usually taking advantage of Coulomb charging, which can force the paired electrons to separate on demand~\cite{Lesovik2001,Recher2001,Hofstetter2009,Herrmann2010,Hofstetter2011,Schindele2012,Herrmann2012,Das2012,Fulop2014,Tan2015,Fulop2015,Borzenets2016,Bruhat2018,Baba2018,Tan2021,Ranni2021,Pandey2021,Brange2021,Scherubl2022,Kurtoessy2022,Ranni2022,Bordoloi2022,Wang2022,deJong2023,Wang2023,Bordin2023,vilkelis2024fermionic,Brange2024}. 

\begin{figure}[!h]
  \includegraphics[scale=0.3]{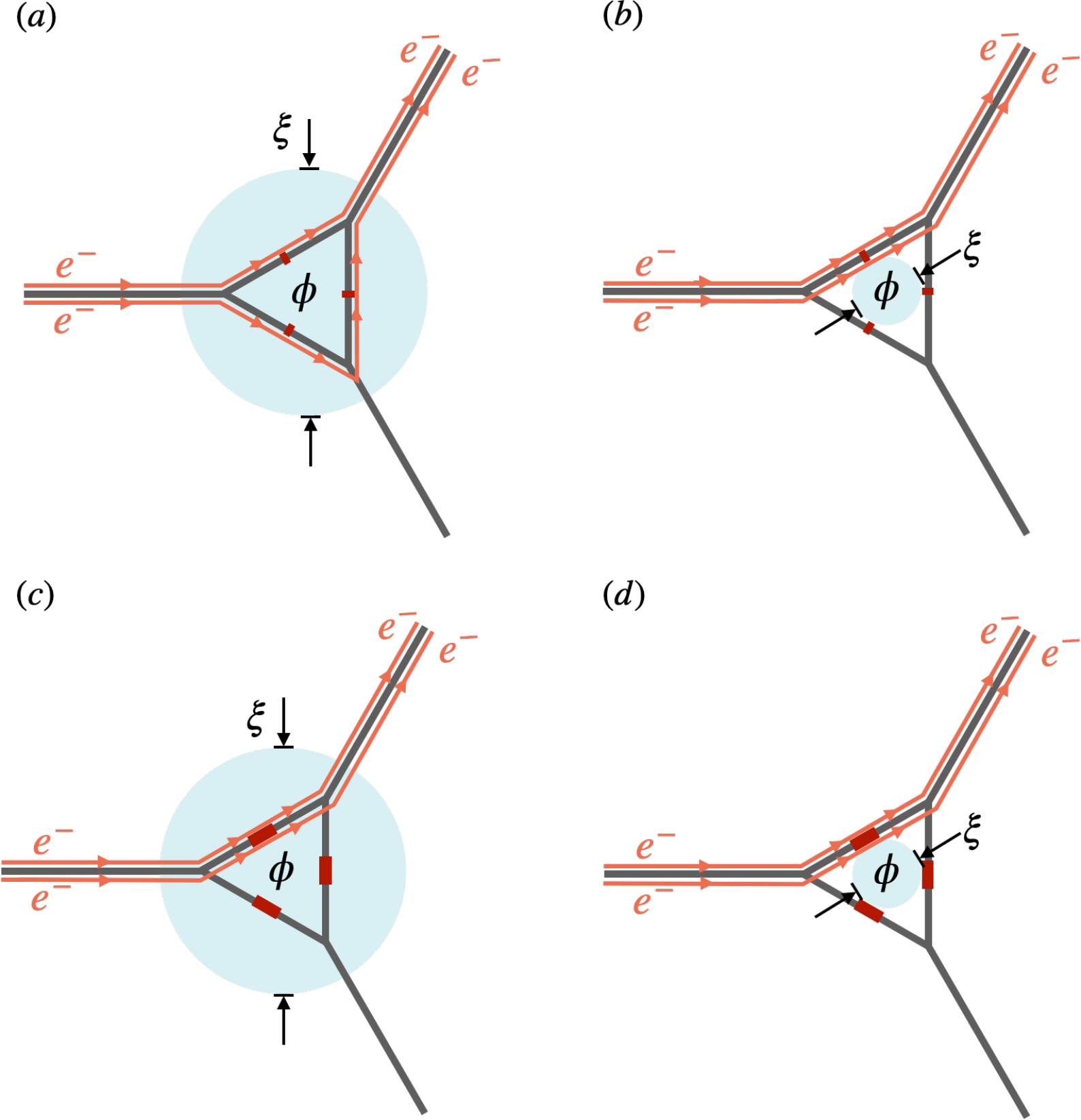}
  \caption{The Y-splitter, the superconducting network component introduced in this work, in two sizes, small (a,c) and large (b,d) compared to the coherence length, $\xi$, and two Josephson coupling strengths, strong (a,b) and weak (c,d). Blue circle represents the coherence length, red segments the insulating barrier thickness. Dominant contribution of Cooper pair transport from the left to the upper arm shown as orange trajectories. The flux $\phi$ through the triangular loop controls interference of single electrons and Cooper pairs. Applying 1/2 (mod 1) flux quanta through the triangular loop leads to destructive interference of Cooper pairs but not single electrons. 
  }
  \label{fig:Y-splitters}
\end{figure}

In this paper, we expand the set of superconducting circuit elements beyond capacitors, inductors, and Josephson junctions, to include a flux-controlled three-terminal Cooper-pair splitter.  We find that including this new component within periodic arrays leads to rich behavior, such as the appearance of flat bands and nontrivial gapped topological phases. Rather than using Coulomb blockade to separate paired electrons, three-terminal splitters contain a loop (like in Refs.~\cite{Chamon-etal-2003,Oshikawa-etal-2006}), and block $2e$ transport by interference, similar to a two-terminal superconducting quantum interference device (SQUID). At the same time, single electrons, whose phase is half as sensitive to flux as Cooper pairs, do not destructively interfere, and so can traverse the loop and recombine at the exits. Note that this looped device, which we denote as a Y-splitter, differs from the usual Cooper pair splitter, where separated electrons typically enter normal leads with or without quantum dots and are not recombined. The separation and recombination of split Cooper pairs from one lead into alternative other leads, similar to a flux-dependent circulator \cite{Chamon-etal-2003, Oshikawa-etal-2006}, steers supercurrent transport.  


We study networks of Y-splitters, and justify analyzing the system within a fermionic language. The “fractionalization” of the Cooper pair into fermions
can be viewed as a way to construct artificial lattices using superconducting wire networks – a solid state alternative to optical lattices. The theoretical analysis of these
fermionic networks is simpler than the study of arrays in terms of Cooper pairs: the fermionic system can be
modeled within the Bogoliubov-de Gennes formalism, while the non-linearities due to the Josephson junctions
render the bosonic-pair system interacting. Specifically, we consider a hexagonal array of superconducting wires with splitters at each vertex. We examine its ground-state properties as a function of flux, identifying several chiral topological superconducting phases characterized by nontrivial Chern numbers. 

The rest of the paper is organized as follows. First, we introduce the flux-tuned three-terminal Y-splitter as a circuit element. Second, we consider an  array of Y-splitters. Third, we derive the phase diagram for the array. Finally, we justify the simple fermionic model by considering a more detailed model.

{\it The single Y-splitter:}
The properties of the individual Y-splitter depends on the size of the loop relative to the superconducting coherence length $\xi$, 
as well as the Josephson coupling strength of its three junctions (assumed to be equal).
Four cases are represented in Fig.~\ref{fig:Y-splitters}, small (a, c) versus large (b, d) loops, and small (c, d) versus large (a, b) Josephson couplings. 
In case (a), where the loop is small compared to $\xi$ and couplings are strong (indicated by thin red barriers), Cooper pairs split then recombine while retaining phase coherence, with the phase difference of the separated electrons, accumulated at the junctions, controlled by flux, $\phi$, through the loop, periodic in the single-electron flux quantum, $hc/e$. 
Cooper pairs can also travel together along each arm, obeying the usual Josephson relations at the junctions within each arm.  In case (b), with a large loop compared to $\xi$, coherent pair splitting and recombining is exponentially suppressed. 
In case (c), coherent splitting is allowed by the small loop but is suppressed compared to pair transport by the weak junctions [thicker red junctions in (c,d)]. 
This is because transport along a single edge is a second order tunneling process, while transferring two electrons along different paths is a third order process. 
For case (d), the large loop and weak tunneling doubly suppress coherent splitting. 
These qualitative descriptions are supported by numerical evidence, as extensively analyzed in Appendix \ref{single_circulator}. There, we examine the behavior of the non-local conductance of a single Y-splitter as a function of the hopping and superconducting parameters. Our results reveal a natural separation into two regimes, which we interpret as indicative of the presence or absence of Cooper-pair splitting. These regimes are in accordance with the limit cases present in Fig. \ref{fig:Y-splitters}.

\begin{figure*}[!ht]
  \centering
  \includegraphics[width=0.8\linewidth]{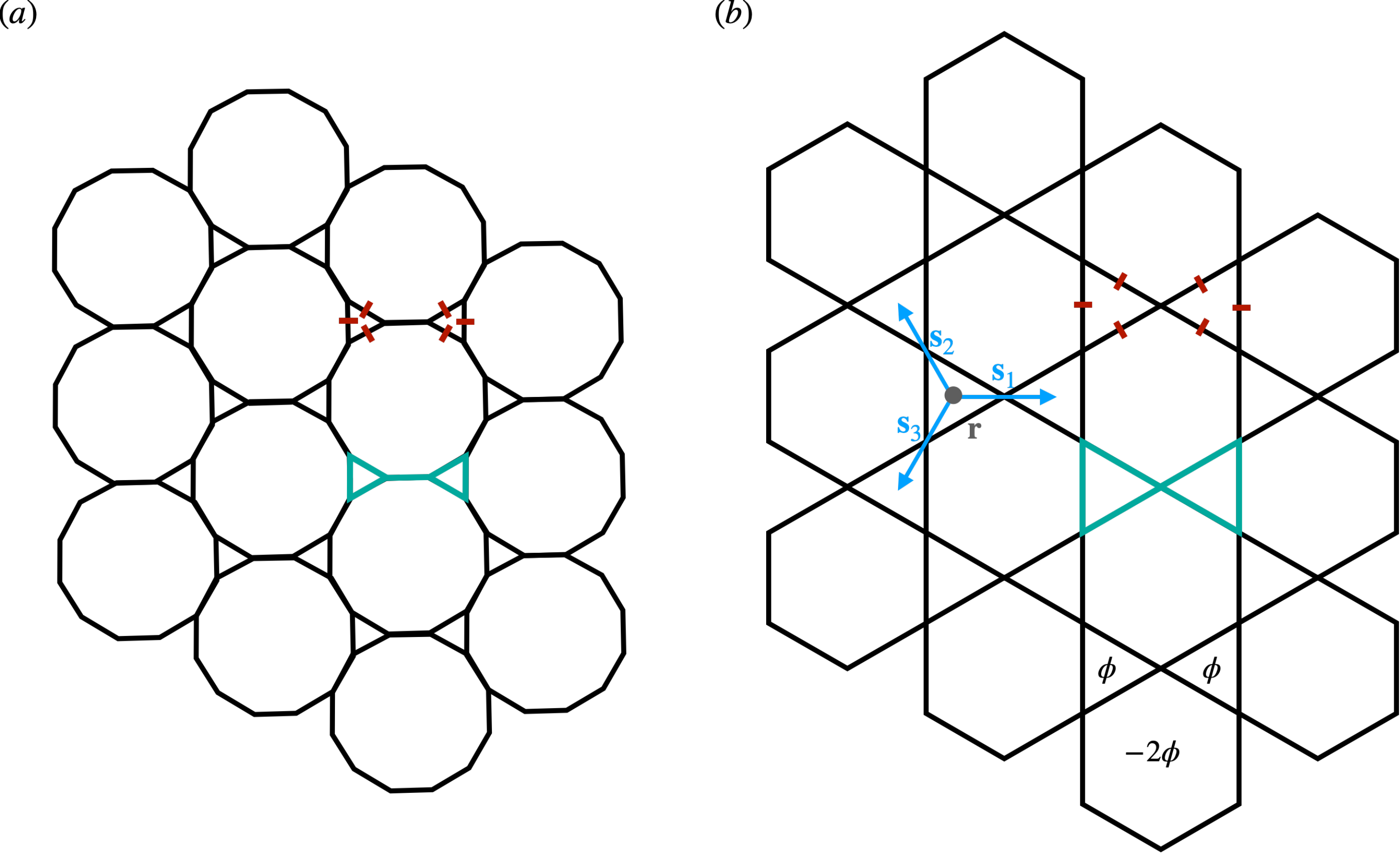}
  \caption{(a) Network of Y-splitters arranged into an Archimedean (3,$12^2$) or star lattice.  By shrinking legs that connect a pair of Y-splitters (once instance is highlighted in green) to a point, we arrive at the kagome lattice shown in (b), with corresponding highlighted green region. (b) Magnetic flux $\phi$ is staggered such that the net magnetic field is zero (mod 2$\pi$). Josephson junctions are located along segments of each triangle, represented in red. The vectors $\mathbf s_i$ (depicted in blue) locate the three nearest left-pointing triangles relative to a lattice site $\mathbf{r}$ at the center of right-pointing triangles.}
\label{fig:kagome_lattice}
\end{figure*}

{\it The Y-splitter network:} Next, we construct a hexagonal 2D array of Y-splitters, which is equivalent to an Archimedean (3,$12^2$) or ``star" lattice of wires, as depicted in Fig.~\ref{fig:kagome_lattice}(a). Notice that as the length of the segment connecting neighboring junctions is shrunk to zero and the two connected Y-junctions become an X-shaped wire crossing, the network morphs into a kagome lattice shown in Fig.~\ref{fig:kagome_lattice}(b). The kagome geometry lead to topological superconductivity (see, e.g., Ref.~\cite{Iimura2018}) in tight-binding fermionic models; by breaking the Cooper pairs with the Y-splitters, we access the fermionic degrees of freedom in an artificially built lattice.

In the kagome configuration [Fig.~\ref{fig:kagome_lattice}(b)], the unit cell area $A_{\rm cell} = A_\triangleright+A_\triangleleft+A_{\varhexagon}$ comprises two triangles of area $A_\triangleright=A_\triangleleft$ each and an hexagon of area $A_{\varhexagon}=6\,A_{\triangleright}$.
We consider applied perpendicular magnetic fields such that the flux per unit cell is an integer multiple of the flux quantum $\Phi_0 = \frac{hc}{e}$, a point to which we return when discussing the experimental considerations for such systems. In this situation, we can study model Hamiltonians that do not demand the use of magnetic super cells (in contrast to Hofstadter-like models. This case of zero net flux (mod one flux quantum) per unit cell is equivalent to flux $\phi$ per triangle and $-2\phi$ per hexagon. (We henceforth work in units in which the flux $\phi$ is an angular variable, $\phi = 2\pi \Phi/\Phi_0$, where $\Phi$ is the physical flux through the triangles, pointing into the page.)

{\it Effective model}: 
First, we analyze an effective model where we replace wires crossing at the sites of the kagome lattice by two fermionic degrees of freedom, or spins. The effective model applies when the superconducting wires in the X-shaped crosses are smaller than or comparable to the superconducting coherence length, $\xi$, a regime in which we expect the bosonic Cooper pairs to split into their fermionic constituents [Fig.~\ref{fig:Y-splitters}(a)]. We shall analyze this effective model first as it is simpler and highlights the physics behind the mechanism to reach the chiral topological superconductivity regime. Then, we will turn to a more detailed tight-binding model in which multiple sites represent the degrees of freedom in each X-shaped cross centered around the vertices of the kagome lattice. The analysis of the more detailed model justifies our use of the simpler effective model.

The centers of the triangles of the kagome lattice form a honeycomb lattice. Let $\mathbf{r}$ denote the sites of the triangular sublattice of this honeycomb lattice corresponding to the right-pointing triangles, which are connected to the left-pointing triangles by vectors 
$\mathbf{s}_1=(1,0)$, $\mathbf{s}_2=(- 1/2, \sqrt{3}/2)$, and $\mathbf{s}_3=(-1/2, -\sqrt{3}/2)$, as shown in Fig. \ref{fig:kagome_lattice}. We work in the Bogoliubov-de Gennes (BdG) basis, which in momentum space is defined by the fields $\Psi_{\mathbf{k}} = [\psi^{}_{\mathbf{k},\uparrow}\;, \psi^\dagger_{-\mathbf{k}, \downarrow}]^{\scriptscriptstyle T }$, with $\psi_{\mathbf{k},\sigma} = [c_{\mathbf{k}, 1, \sigma}\;, c_{\mathbf{k}, 2, \sigma}\;, c_{\mathbf{k}, 3, \sigma}]^{\scriptscriptstyle T }$, where $c_{\mathbf{k}, a, \sigma}$ is the electron annihilation operator with momentum $\mathbf{k}$, at sublattice $a=1,2,3$, and spin $\sigma=\uparrow, \downarrow$.

The matrix Hamiltonian for the effective model is
\begin{subequations}
\label{eq:effective-H}
\begin{align}
H_{\mathbf{k}} = 
\begin{bmatrix}
    h_{\mathbf{k}} & \Delta_{\mathbf{k}}\\
    \Delta^*_{\mathbf{k}} & -h^*_{-\mathbf{k}} 
\end{bmatrix}
\;,
\end{align}
where
\begin{align}
\label{H_Kagome}
h_{\mathbf{k}} = 
\begin{bmatrix}
    0 & \alpha^{\,}_{\triangleright} + \alpha^{\,}_{\triangleleft}\,d_{\mathbf{k}}^{12} & \bar\alpha^{\phantom\dagger}_{\triangleright} + \bar\alpha^{\phantom\dagger}_{\triangleleft}\,d_{\mathbf{k}}^{13} \\
    \bar\alpha^{\phantom\dagger}_{\triangleright} + \bar\alpha^{\phantom\dagger}_{\triangleleft}\,d_{\mathbf{k}}^{21} & 0 & \alpha^{\,}_{\triangleright} + \alpha^{\,}_{\triangleleft}\,d_{\mathbf{k}}^{23} \\
    \alpha^{\,}_{\triangleright} + \alpha^{\,}_{\triangleleft}\,d_{\mathbf{k}}^{31} & \bar\alpha^{\phantom\dagger}_{\triangleright} + \bar\alpha^{\phantom\dagger}_{\triangleleft}\,d_{\mathbf{k}}^{32} & 0
\end{bmatrix}
\end{align}
and
\begin{align}
\Delta_{\mathbf{k}} =
  \Delta\;
  \begin{bmatrix}
    e^{i\theta_1} & 0 & 0\\
    0  & e^{i\theta_2} &0\\
    0 &0 & e^{i\theta_3}
  \end{bmatrix}
\;, 
\end{align}
\end{subequations}
with $d_{\mathbf{k}}^{ij}= e^{-i\mathbf{k}\cdot (\mathbf{s}_i-\mathbf{s}_j)}$ and 
$\alpha_{\triangleright}=\alpha_{\triangleleft}= \Gamma\;e^{i\phi/3}$. $\Gamma$ is the amplitude for fermion tunneling between sites and $\Delta e^{i\theta_a}$ is the on-site s-wave BCS order parameter associated to X-crosses. We allow for three independent superconducting phases $\theta_a$, $a=1,2,3$, on the three sublattices. These superconducting phases are determined self-consistently, by minimizing the total many-body ground state energy, which is obtained by filling the negative energy single-particle eigenstates of the Hamiltonian Eq.~\eqref{eq:effective-H}. Numerically, we sweep for values of $\theta_2-\theta_1$ and $\theta_3-\theta_1$ that minimize the ground state energy (gauge freedom allows one to set $\theta_1 = 0$).
\begin{figure*}[!ht]
	\centering
\includegraphics[width=1.0\linewidth]{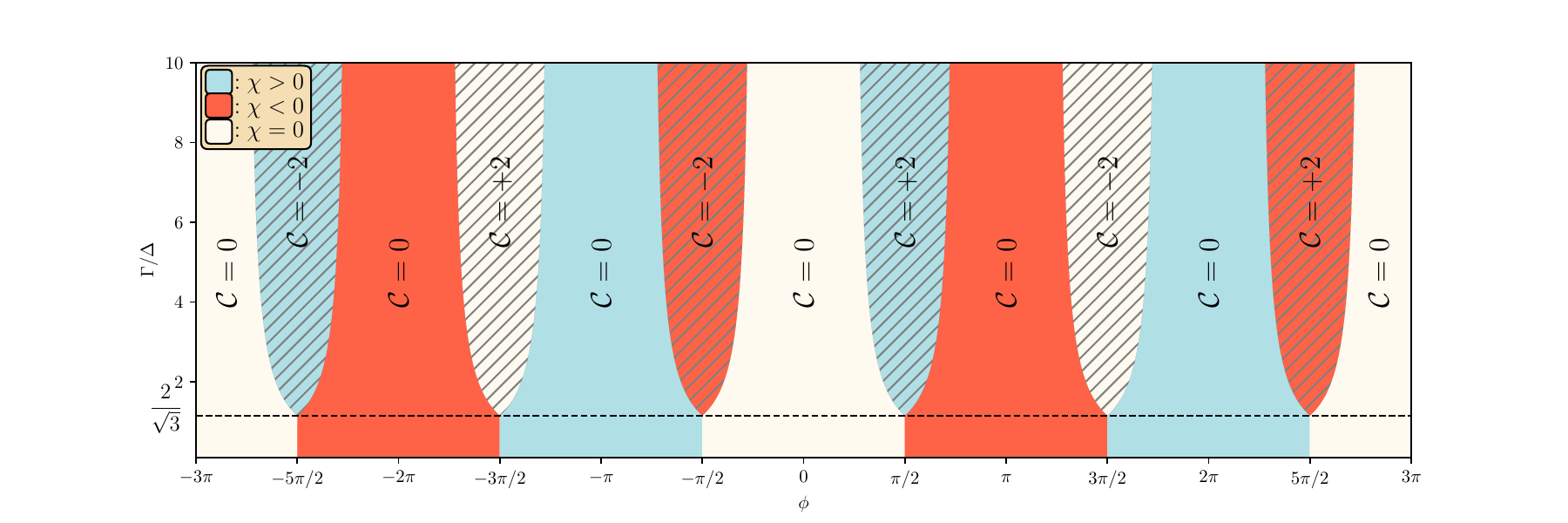}
	\caption{Phase diagram of the effective model. The diagram shows the chirality $\chi$ and Chern numbers $\cal{C}$
          as functions of the flux $\phi$ and ratio
          $\Gamma/\Delta$. The chirality takes specific values $\chi_{23}=0,\pm 1$ per Eq.~\eqref{eq:chirality}, denoted as $\chi=0, >0, <0$ with different colors. The limit $\Gamma\ll \Delta$ is the ``classical'' Josephson regime, which is studied in more detail in Appendix~\ref{classical}. Regions with non-zero Chern numbers (hashed) only appear in the regime in which Cooper-pair splitting is active.}
	\label{fig:orderpar}
\end{figure*}
Let us define the following quantities
\begin{eqnarray}
  \label{eq:chirality}
  \chi_{ab} = \frac{2}{\sqrt{3}}\;\sin(\theta_a - \theta_b)
  \;,
\end{eqnarray}
 that we refer to as \textit{chiralities}. They are proportionally related to the Josephson current, $I_J = I_c\sqrt{3}/2 \, \chi$, flowing along the kagome wire network, and measure the superconducting phase differences between sublattices.  Josephson currents do not flow for $\chi_{ab}=0$, while currents flow maximally clockwise or counter-clockwise when $\chi_{12} = \chi_{23} = \chi_{31}=\mp 1$, respectively. {These equalities suggest that $\mathbb Z_3$ rotation symmetry is preserved in the low-energy physics of the system.} In Fig. \ref{fig:orderpar}, we show $\chi\equiv\chi_{12}$ 
as function of the flux $\phi$ and the ratio $\Gamma/\Delta$, which displays a repeating pattern of values $\chi = 0,\pm 1$.


The regime $\Delta\gg \Gamma$ corresponds to the limit where Cooper pair splitting is suppressed [Fig.~\ref{fig:Y-splitters}(c) and (d)]; in this case, the superconducting phase relations conform to those expected from Josephson junctions with $E_J\sim \Gamma^2/\Delta$, as we discuss in detail in the Appendix~\ref{classical}. There we also discuss the emergence of periodicity $6\pi$ (or $3\,\Phi_0$) in the dependence in $\phi$ present in Fig.~\ref{fig:orderpar} (instead of periodicity $2\pi$).

We next focus on the Chern numbers associated with the filled negative energy single-particle states of Hamiltonian Eq.~\eqref{eq:effective-H}. {The Chern numbers count the number of monopoles in momentum space present in the many-body ground state, resulting in the quantized topological invariants 
\begin{eqnarray}
    \mathcal C =\frac{1}{2\pi}\sum_{\text{filled bands}} \int_{\operatorname{BZ}} d^2 \mathbf{k}  \, \left[\partial_{k_x} A_y - \partial_{k_y} A_x\right],
\end{eqnarray}
where the integral is performed in the first Brillouin zone $\operatorname{BZ}$, and $A_i = \bra{n(\mathbf k)} \partial_{k_i} \ket{n(\mathbf k)}$ is the Berry connection. In the presence of boundaries, the sign of $\mathcal{C}$ determines the direction of the edge currents: they circulate counterclockwise for $\mathcal{C} > 0$ and clockwise for $\mathcal{C} < 0$.}
Fig.~\ref{fig:orderpar} shows these Chern numbers ${\cal C}$, as function of the flux $\phi$ and ratio $\Gamma/\Delta$, indicating trivial and topological phases. The regions in the phase diagram where
${\cal C}=\pm 2$ coincide with those for which the chirality $\chi$ departs from the classical Josephson-array behavior, i.e., those regions where Cooper pair splitting is effective, for $\Gamma$ dominating over $\Delta$. In fact, the dome-like regions have their tip at the ``magic" values $\phi=\pm \pi/2, \pm 3\pi/2, \pm 5\pi/2$, flux values for which Cooper pairs interfere destructively going around the triangles of the lattice, since they accrue phases $2\phi=\pm \pi, \pm 3\pi, \pm 5\pi$. In contrast, the fermions (fractions of the pair) {\it do not} destructively interfere; such  values of flux maximally break time-reversal symmetry.

\begin{widetext}

\begin{figure}[ht]
    \centering
    \includegraphics[scale=0.3]{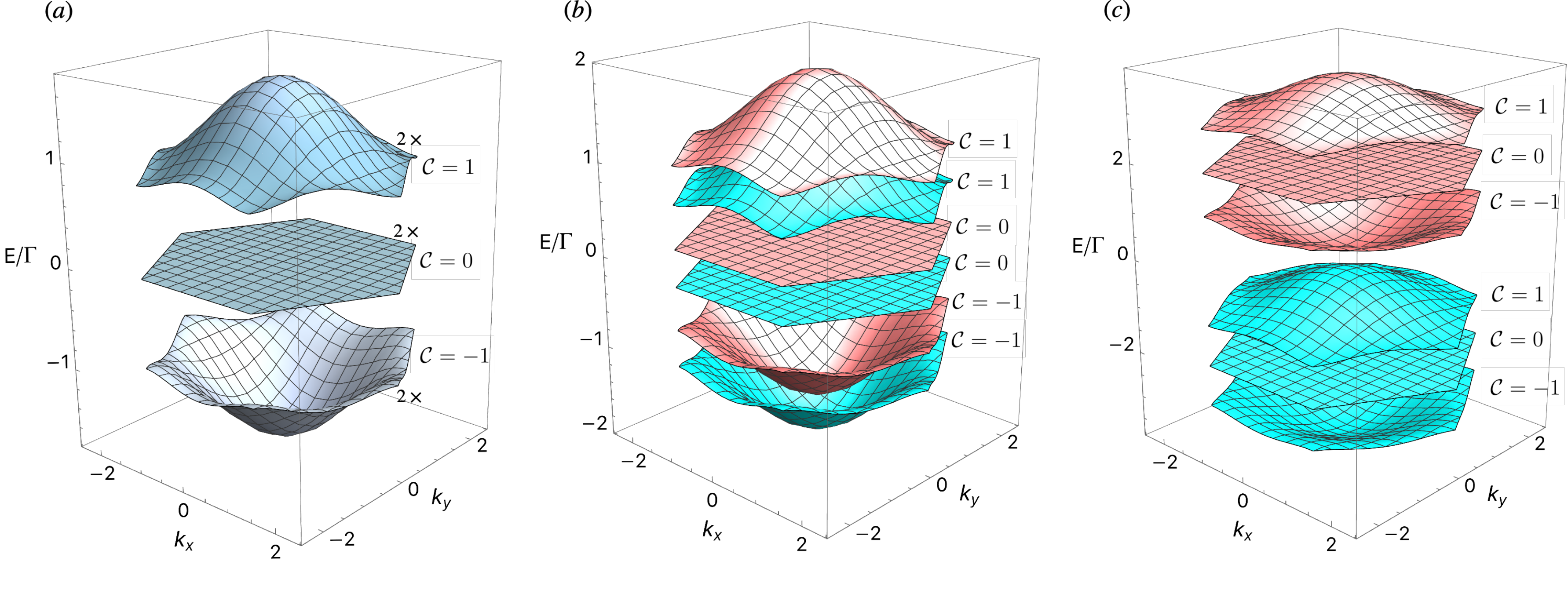}
     \caption{Bands in the flux configuration $\phi=3\pi/2$ for (a) $\Delta=0$, where each one of the three bands is double (spin) degenerate;  (b) $0<\Delta< \sqrt{3}\, \Gamma/2$, where the degeneracy in each one of the bands is split by $\pm \Delta$ as in Eq.~\eqref{eq:spectrum-phi-equal-pi-half}; and (c) $\Delta>\sqrt{3}\Gamma/2$, where the bands are separated into two sets of spectrum of $h_{\mathbf k}$. The light blue and pink colored bands in (b) and (c) indicate anti-symmetric and symmetric spin states, which are non- degenerate when $\Delta\neq 0$.
     }
    \label{fig:splitbands}
\end{figure}
\end{widetext}

It is instructive to analyze the special case where $\phi=3\pi/2$, for which $h_{\mathbf{k}}=-h^*_{-\mathbf{k}}$. For this value of flux we found numerically that $\chi=0$ (see Fig~\ref{fig:orderpar}). The Hamiltonian in Eq.~\eqref{eq:effective-H} then simplifies to
\begin{align}
  H_{\mathbf{k}} =
  \mathds{1}_{2\times 2}\otimes h_{\mathbf{k}}+\Delta \, \sigma_{\rm x}\otimes  \mathds{1}_{3\times 3}
  \label{eq:special_flux}
  \;,
\end{align}
where $\sigma_{\rm x}$ is a $2\times 2$ Pauli matrix and $\mathds{1}_{p\times p}$ corresponds to a $p\times p$ identity matrix. The two terms in Eq.~\eqref{eq:special_flux} commute and can be diagonalized separately. The spectrum of $h_{\mathbf{k}}$ is symmetric with respective to zero energy, and thus one of its three eigenvalues is zero for all values of momentum ${\mathbf{k}}$, i.e., it contains a flat band (the model without pairing was studied in Ref.~\cite{Green-Santos-Chamon-Kagome}). The other two eigenvalues of $h_{\mathbf{k}}$ are given by $\pm \varepsilon_{\mathbf k}$ with
\begin{widetext}
\begin{align}\label{band_lobe}
  \varepsilon_{\mathbf k}
  =
  \dfrac{\Gamma}{\sqrt{2}}
  \sqrt{
  3+\cos(\sqrt{3}k_x)
  +
  \cos\left(\dfrac{\sqrt{3}}{2} k_x-\dfrac{3}{2}k_y\right)
  +
  \cos\left(\dfrac{\sqrt{3}}{2} k_x+\dfrac{3}{2}k_y\right)}
  \;.
\end{align}
\end{widetext}
The eigenvalues of the full Hamiltonian $H_{\mathbf{k}}$, in  Eq.~\eqref{eq:special_flux}, are thus
\begin{align}
  \label{eq:spectrum-phi-equal-pi-half}
  \pm \Delta\,,
  \quad
  -\varepsilon_{\mathbf k}\pm\Delta\,,
  \quad
  \varepsilon_{\mathbf k}\pm\Delta\,.
\end{align}
The splitting of bands is shown graphically in Fig.~\ref{fig:splitbands}.
There is a transition point, for $\phi=3\pi/2$ and $\Delta/\Gamma=\sqrt{3}/{2}$, where the middle flat bands at $\pm \Delta$ touch the dispersive bands. For $\Delta/\Gamma<\sqrt{3}/{2}$ the Chern number for the filled negative energy bands equals ${\cal C}=-2$, a regime where the system realizes a chiral topological superconductor. Again we remark that Cooper-pair splitting is maximally enhanced for $\Delta\ll \Gamma$ and $\phi = \pm \pi/2, \pm 3\pi/2, \pm 5\pi/2$, where destructive interference suppresses Cooper pair motion around the kagome triangles. When $\Delta \gg \Gamma$ the Chern numbers at half filling are trivial and no topological phase appears. In this limit, the spectrum corresponds to two well-separated copies of the three bands of $h_{\mathbf{k}}$. The total Chern number at half-filling is the sum of those of $h_{\mathbf{k}}$, which sum to zero.

\begin{figure}[ht]
  \centering
  \includegraphics[width=0.4\linewidth]{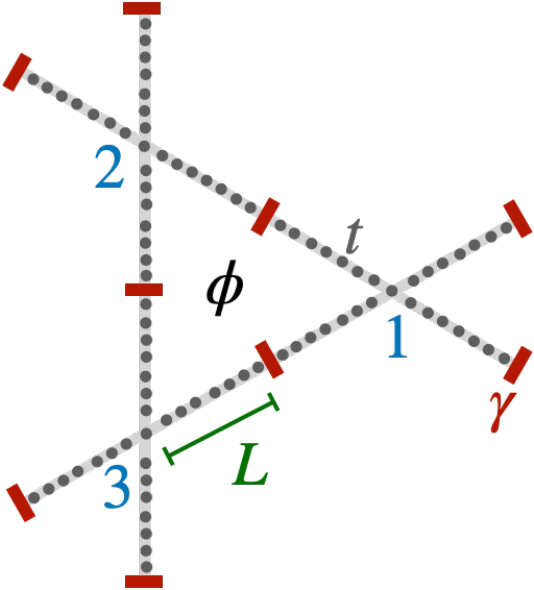}
  \caption{Model of the wire network that accounts for the finite extent of the wires. The model can be understood as a network of X-shaped crossings comprising two intersecting wires at the sites of the kagome lattice. Intra-molecule tight-binding hopping have amplitude $t$, and inter-molecule hopping have amplitude $\gamma$. Intra-molecule superconductivity is accounted for within the BdG formalism. We label the three X-shaped crossings 1, 2, and 3, according to our site labeling convention in Fig.~\ref{fig:kagome_lattice}(b)}
  \label{fig:unitcell}
\end{figure}

{\it Tight-binding model for the superconducting wire network}: 
A more detailed model of the network that captures the finite extent of the wires is depicted in Fig.~\ref{fig:unitcell}. This extended model allows us to justify our previous minimal approach, where each wire was approximated by a single site. The model can be thought of as a network of X-shaped crossed wires, which we refer to as \textit{X-molecules}, formed by two intersecting wires at the sites of the kagome lattice. The intra-molecule Hamiltonian is modeled by a tight-binding Hamiltonian where electrons hop along each wire with amplitude $t$. Similarly to the effective model,  superconductivity is accounted for by an on-site BCS pairing interaction $\Delta$, and inter-molecule hopping amplitude is accounted for by $\gamma$.

\begin{widetext}

\begin{figure}[ht]
  \centering
\includegraphics[scale=0.47]{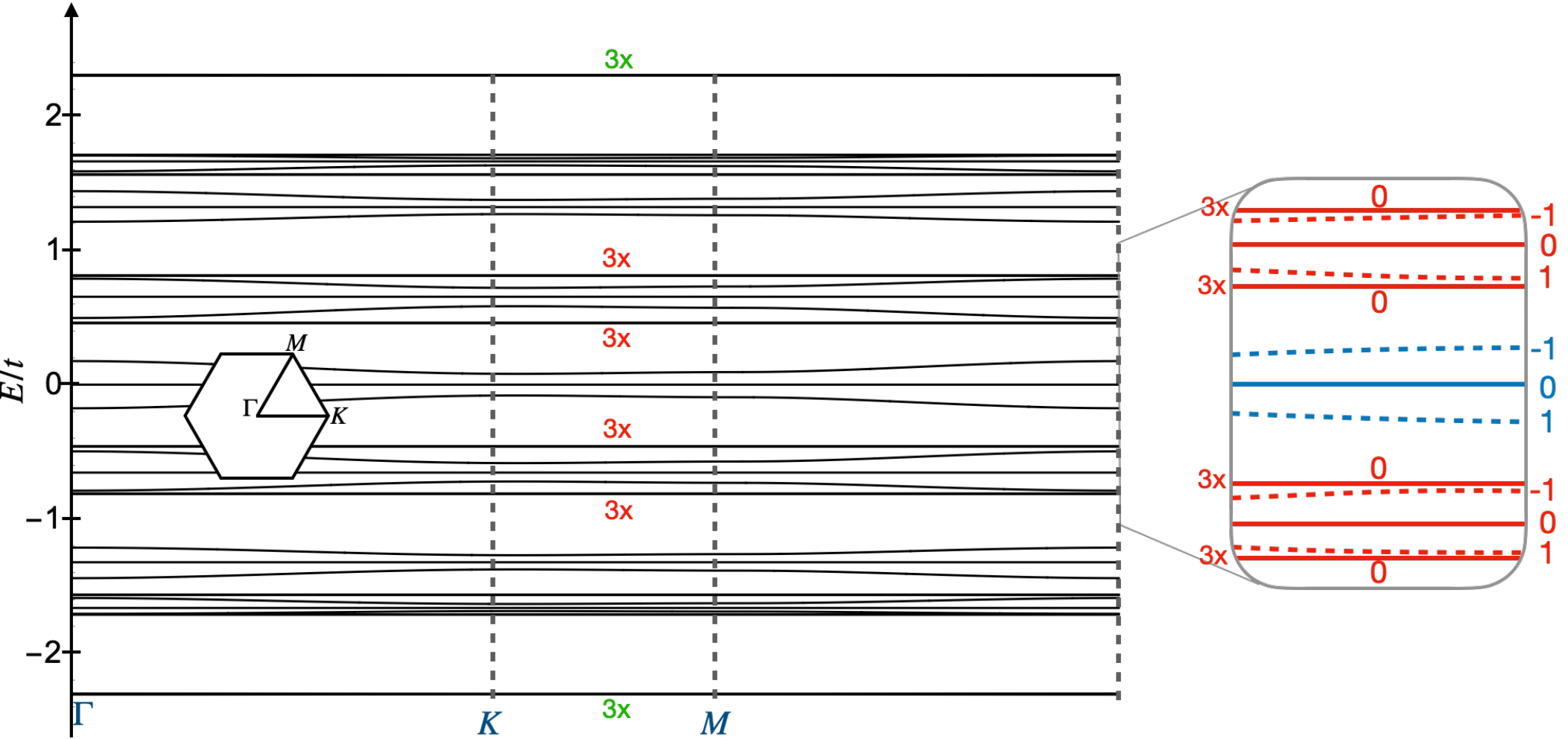}
  \caption{Band structure of the wire network model for $\Delta=0$, $\gamma = t/2$ and ``magic" flux $\phi = \pi/2$. This is a numerical result for $L=4$ along the high-symmetry points $\Gamma-K-M$ in the Brillouin zone. Note the existence of 3x (highlighted in red) degenerate flat bands that sandwich clusters of bands, and are separated from the clusters by a non-zero gap. (These bands are flat for all values of $\phi$.) The inset shows two types of clusters of bands, depicted in blue and red, with the band Chern numbers ($\pm 1,0$) labeled next to each band. The bands in the middle of both blue and red clusters (depicted in full line in the inset), are flat at the magic values of the flux $\phi = (n-1/2)\pi$, with $n\in \mathbb{Z}$. The non-flat bands depicted in dashed lines are the only ones that carry non-trivial Chern-numbers $\pm 1$.}
  \label{fig:moleculebands}
\end{figure}
\end{widetext}

The Appendix \ref{molecule} contains a detailed study of the tight-binding model describing the wire network. Here we summarize the main results. In Fig.~\ref{fig:moleculebands} we show the spectrum of the wire network for the case in which superconductivity is turned off, for fluxes $\phi = (n-1/2)\pi$, with $n\in \mathbb{Z}$. We observe that at half-filling the spectrum of the three bands around zero energy coincides with that of the effective model, with the flat band at zero energy. The Chern numbers of the band below and above the flat band coincide with those of the effective model (and the Chern numbers of the other core filled levels add to zero). Upon turning on superconductivity, the spectrum assumes the form of that in Eq.~\eqref{eq:spectrum-phi-equal-pi-half}. Therefore, the more detailed treatment of the wire network yields the same conclusions as the simpler effective model with fewer bands. We stress that this conclusion requires that the spacing between the clusters of bands in Fig.~\ref{fig:moleculebands} are larger than the superconducting gap $\Delta$. This condition is equivalent to requiring that the length scale of the X-shaped wire crosses is smaller than the superconducting coherence length $\xi$. Recall that we already stated and physically motivated such a condition earlier, when first presenting the effective model.

Before moving on to experimental considerations, we remark on an interesting feature of the tight binding wire network: the clusters of bands are sandwiched between completely flat bands away from zero energy that are gapped for \textit{all values} of $\phi$ (Fig.~\ref{fig:moleculebands}). Such bands have been of interest recently, for example in twisted double-layer graphene \cite{MacDonald-Graphene}. On the other hand the flat band at zero energy is only flat at the magic values of $\phi$.

{\it Experimental considerations}: 
Experimentally, it is most natural to impose a uniform magnetic field on the entire lattice. The condition of zero flux per unit cell of the lattice can still be enforced by taking advantage of the $2\pi$-periodic flux dependence. For the kagome configuration, fluxes threading triangles and hexagons must be equal to, respectively, $\phi+2\pi\,m$ and $-2\phi+2\pi\,n$, with $m,n$ integers. For a uniform field, the ratio of these fluxes equals the ratio of the areas $A_\triangleright/A_{\varhexagon}=1/6$, yielding discrete values of uniform fields that satisfy the zero-total-flux condition 
\begin{align}
  \phi=\frac{\pi}{4}(n-6m)
  \,.
\label{eq:flux_condition2}
\end{align}
Fluxes that do not satisfy Eq.~\eqref{eq:flux_condition2} result in magnetic unit cells larger than the lattice unit cell. Fluxes that do satisfy Eq.~\eqref{eq:flux_condition2},  namely, $\phi=\pm \pi/2, \pm 3\pi/2, \pm 5\pi/2$, etc., maximize the degree of Cooper-pair splitting, and yield topological superconductors above the minimal value $\Gamma/\Delta = 2/\sqrt{3}$. Lattices other than the kagome, such as the example in Fig.~\ref{fig:kagome_lattice}(a), can also yield topological superconductors at a discrete set of uniform fields, using the length of the horizontal links to tune the ratio $A_\triangleright/A_{\varhexagon}$, where the denominator would now be the area of the dodecagons rather than hexagons.

A challenge for experimental implementation is creating Y-splitters that are comparable in size to the superconducting coherence length $\xi$. 
A good choice of material is aluminum with its long bulk coherence length,  $\xi_0 = \hbar v_{F}/\pi \Delta \sim 1.6\,\mu$m, which appears to readily allow structures smaller than $\xi$ with reasonable magnetic field scales (recalling that $hc/2e \sim 2$ mT$\,\mu$m$^2$). However, in thin, narrow, disordered, or granular aluminum, the coherence length is considerably reduced, typically to tens or to hundreds of nanometers, in deposited thin films (roughly given by the geometric mean of $\xi_0$ and film thickness) and nanowires \cite{Morgan-wall2015}. 

Two routes to extending the superconducting coherence length, allowing realistic Y-splitter dimensions and magnetic field scales (below $\sim$ 0.1~T) can be pursued. For metallic structures using traditional Josephson junctions, careful attention to film morphology, via deposition and annealing, can reduce granularity and yield coherence lengths of order 1$\,\mu$m in films \cite{Mayadas1968, Mayadas1969} and wires \cite{Lee2010}. 
Alternatively, epitaxial semiconductor-superconductor heterostructures can take advantage of the proximitized high-mobility semiconductor to extend coherence length to micron scales \cite{Mayer2019}. Previous studies of coherence of Andreev bound states in lithographically patterned 2D hybrid Al/InAs heterostructures showed coherence lengths order 1 $\mu$m \cite{Poeschl2022, Whiticar2020}. Length dependence studies specific to Cooper pair splitters have not been reported to our knowledge. Splitters mediated by integrated quantum dots have been realized in superconductor-semiconductor hybrid nanowires \cite{Hofstetter2009, Hofstetter2011, Das2012} and 2D heterostructures \cite{Wang2023}. In those cases, the superconductor width was comparable to the bare Al superconducting coherence length, so any enhancement of coherence length from the underlying high-mobility semiconductor was not explored.

Optimizing Cooper pair splitting also requires balanced junctions, that is, three equal junctions in each Y-splitter. Within hybrid materials, each of the three junctions in all Y-junctions across the array can be electrostatically tuned simultaneously using three independent gates, separated by insulating layers \cite{Banszerus2024}. Local variation among Y-splitters cannot be readily corrected this way. We have not considered the role of non-uniformity of junctions across the array on the phase diagram but expect reasonable robustness of the phases, analogous to parallel coupling of topological and trivial superconducting states. 

Regarding experimental tests to validate the predicted phase diagram, we propose measuring thermal Hall voltage, which is expected to be quantized in proportion to the Chern number as a function of flux and gate-controlled Josephson coupling strength \cite{Sumiyoshi2013, Iimura2018, Glodzik2023}. This could follow related experiments in naturally occurring materials~\cite{Kasahara2018}. Other signatures, such as non-sinusoidal current-phase relations \cite{Kurter2015} in Josephson junctions could be realized as a gated discontinuity in the array. Other bulk 2D signatures \cite{Li2015} could also be investigated.

While only discrete fluxes compatible with Eq.~\eqref{eq:flux_condition2} are expected to yield topological superconducting phases, this condition still allows access to all phases, so that flux and gate voltages can be used as independent control parameters. We note that thermoelectric measurements have also been investigated theoretically \cite{Cao2015} and experimentally \cite{Arora2024} in single Cooper pair splitters with blockading quantum though the connection to bulk thermal Hall effect in an array of Y-splitters is not clear. Alternative measurements 

{\it Discussion and broader perspectives}: Arrays of Y-splitters considered in this paper are but one concrete example of a broader class of superconducting circuits in which Cooper pairs are intentionally fractionalized as means of experimentally realizing effective lattice models of fermions, a solid-state electronic counterpart to the artificial lattices built in atomic and optical systems. The key element that brings the superconducting array to the regime in which the system can be analyzed as a fermionic model is that the lattice spacing is smaller than (or of the order of) the coherence length. It is this interplay of scales that brings the network to the quantum regime in which phases such as chiral topological superconductivity can emerge.

In these superconducting circuits, the application of the magnetic field serves as a parameter to tune the electronic band structure. By constraining the values of fields to those that yield zero net flux per unit cell, we can analyze the fermionic spectrum in simple terms. In this paper we identify specific discrete values of magnetic flux for which the spectrum contains flat bands and bands with non-trivial Chern numbers. These fluxes coincide with the values that yield destructive interference for Cooper pairs going around the triangles of the kagome lattice, and thus these values also maximize Cooper-pair splitting.

These results indicate that superconducting arrays that include phase-controlled components that are small or comparable in size to the superconducting coherence length provide a novel resource for creating superconducting metamaterials with quantum phases that are seemingly not otherwise available. Understanding such metamaterials goes beyond a bosonic description of superconducting arrays, including quantum arrays with Josephson and charging energy, and are best addressed by fermionic models. 

One can naturally extend the present results for arrays to non-periodic systems of granular matter with small structures. Of course, there is a trade-off: components must be smaller than the superconducting coherence length but still allow interference to be controlled by flux. In this regime, it appears that periodicity is not required. This further opens new avenues to engineer and explore exotic quantum phases.

\vspace{.5cm}
\begin{acknowledgments}
This work is supported in part by the NSF Grant DMR-1906325 (C.C.~and G.D.), the ERC Synergy Program (C.M.M.), the Villum Young Investigator Program (S.V.), and a research grant (Project 43951) from VILLUM FONDEN (C.M.M.~and S.V.).
C.C. thanks the hospitality of the Thouless Institute for Quantum Matter at the University of Washington.
\end{acknowledgments}

\appendix


\onecolumngrid
\setcounter{secnumdepth}{2}

\renewcommand{\appendixname}{Appendix}

\appendix

\section{Transport properties of a single Y-splitter}\label{single_circulator}

In this section we investigate the physics of a single Y-splitter using a tight-binding model for superconducting wires, implemented numerically in \textit{Kwant}, an open-source Python package \cite{groth2014kwant}. The results offer quantitative insight into the physics of  Y-splitters, which is further enhanced through a link between non-local conductance matrix elements and Cooper-pair splitting.

\begin{figure}[h]
	\centering
	\includegraphics[width=0.5\linewidth]{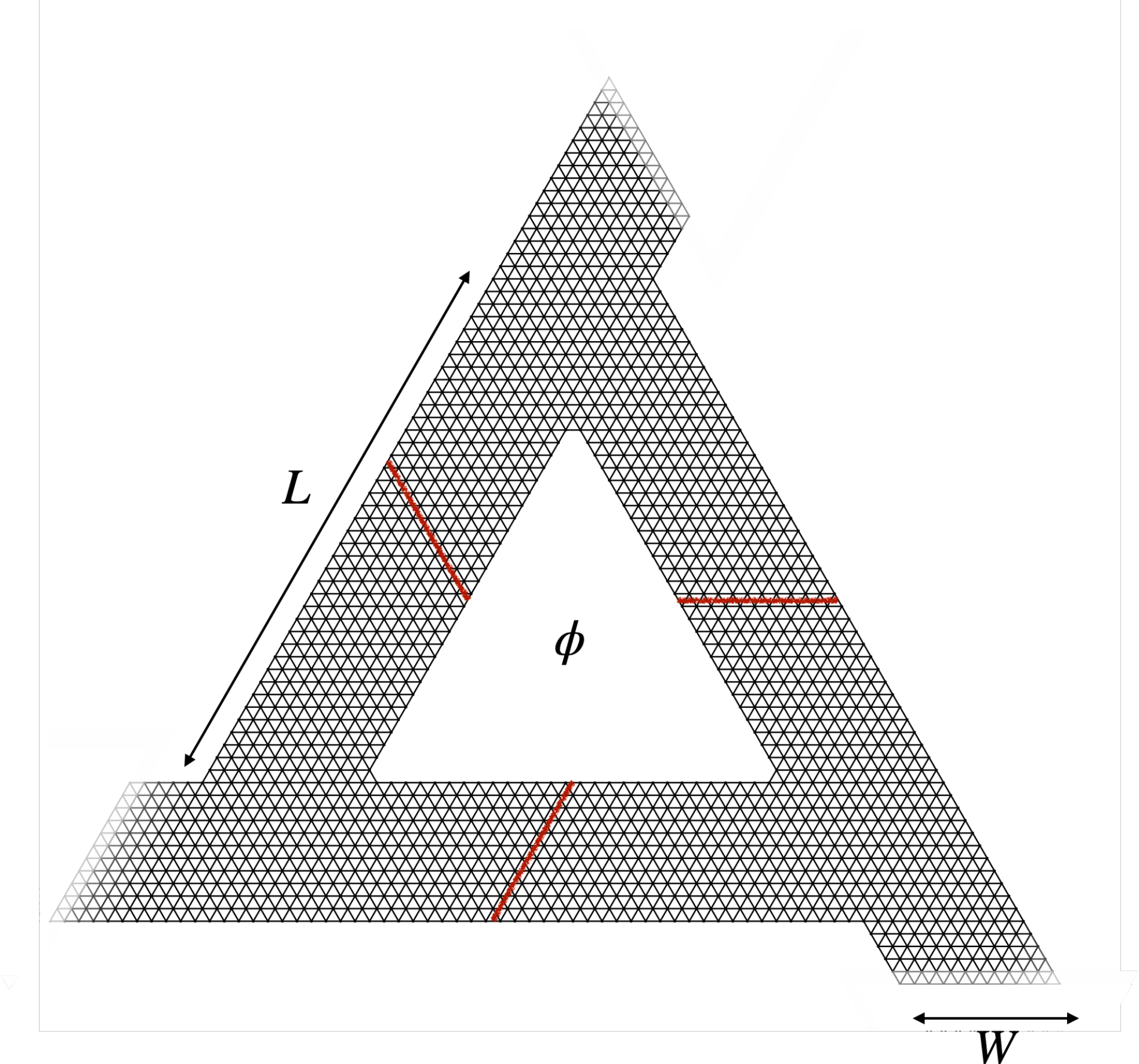}
	\caption{Triangular lattice modeling of a single Y-splitter with flux $\phi$ attached to three leads. Red lines separate two wires and indicate Josephson junctions.}
	\label{circulator_leads}
\end{figure}

The system is defined on a two-dimensional triangular lattice with lattice spacing \( a \), inscribed within an equilateral triangle of edge length \( L \) and width \( W \), as illustrated in Fig. \ref{circulator_leads}.  
Three infinitely long normal leads, each of width \( W \), are attached at the triangle's vertices, preserving \(\mathbb{Z}_3\) rotational symmetry. 
{Each edge of the triangle is separated into two halves that are connected by Josephson barriers (red lines in Fig. \ref{circulator_leads}). 
Electrons hop across the junctions with amplitude $\gamma$, acquiring a Peierls phase $e^{i\phi/3} $ when moving counterclockwise.  
Within the same region (other than the barrier)  they hop with strength $t$, while an on-site superconducting potential \( \Delta \) pairs spin-up and spin-down electrons.} 

We expect to observe Andreev reflection at the normal-superconductor (NS) interface, where an incident electron reflects as a hole. In a multi-terminal setup, crossed Andreev reflection (CAR) can also occur, in which the reflected hole emerges from a different lead than the one through which the electron entered, requiring the measurement of non-local conductance components. As noted in previous studies \cite{wang2011interference, deacon2015cooper}, CAR can play a crucial role in detecting Cooper-pair splitting.

Let \( G_{ij} \) denote the conductance associated with the transmission probability $T_{ij}$ that an electron injected the $i$-th lead travels to the $j$-th one
\begin{eqnarray}
    G_{ij} = \frac{e^2}{h}T_{ij}.
\end{eqnarray}
The diagonal elements of the conductance matrix, \( G_{ii} \), can be determined through local measurements.
 More relevant to our study are the non-local conductance elements \( G_{ij} \) with \( i \neq j \), which serve as indicators of Cooper-pair splitting by probing non-local correlations in scattering processes. These elements are also closely linked to crossed Andreev reflection (CAR). 
Specifically, we define:  
\[
G_A \equiv \frac{1}{3} \epsilon_{ij} G_{ij},
\]
where \( \epsilon_{12} = \epsilon_{23} = \epsilon_{31} = 1 \), \( \epsilon_{21} = \epsilon_{32} = \epsilon_{13} = -1 \), and zero otherwise. The conductance \( G_A \) quantifies transport asymmetry, favoring a specific spatial direction dictated by the magnetic flux. We expect \( G_A \) as a function of \( \phi \), \( \Delta \), and \( \gamma \) to serve as an indicator of the dominant charge carriers in different regimes: for values of \( \phi \) that are odd-multiples of $\pi/2$ and sufficiently large correlation lengths, transport is dominated by single electrons, whereas in other cases by Cooper pairs.

\begin{figure}[ht]
	\centering
	\includegraphics[width=0.7\linewidth]{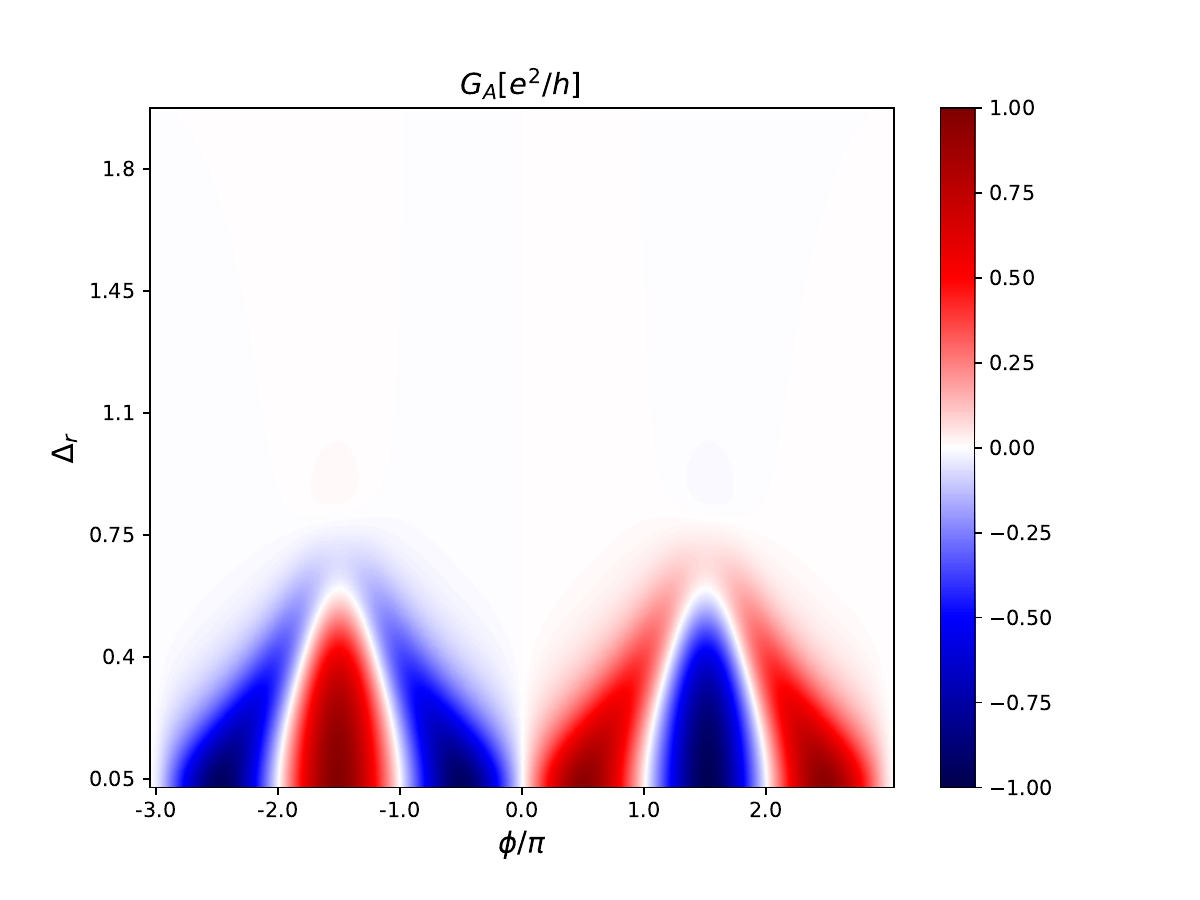}
	\caption{Heatmap of $G_A$ as a function of the flux $\phi$ and the ratio ${\Delta_r}$.}
	\label{GA_heatmap}
\end{figure}

We define the dimensionless ratios  
\[
{\gamma}_r = \gamma/t, \quad {\Delta}_r = \Delta/t, \quad {L}_r = L/a,
\]  
where energy and length scales are expressed in units of \( t \) and \( a \), respectively. This formulation is advantageous as it allows us to focus solely on the dependence of \( G_A \) on the three parameters, together with magetic flux.  Fig. \ref{GA_heatmap} presents \( G_A \) (in units of \( e^2/h \)) as a function of the magnetic flux \( \phi \) and the superconducting pairing strength \( {\Delta}_r \), for a fixed system size \( {L}_r \) and coupling parameter \( {\gamma}_r \).

The magnitude of \( {\Delta}_r \) suggests the existence of two distinct regimes in the system, with \( G_A \) serving as a non-local order parameter: it takes nontrivial values for \( {\Delta}_r \ll 1 \) and vanishes for \( {\Delta}_r \gg 1 \).  
The system's behavior in these two regimes is qualitatively different and can be understood in their respective asymptotic limits. For large \( {\Delta}_r \), the system behaves as a conventional \( s \)-wave superconductor carrying supercurrent, whose conductance does not display an asymmetry that is captured by \( G_A \). In contrast, for small \( {\Delta}_r \), the physics remains adiabatically connected to that of free electrons, where conductance asymmetry reaches its maximum \cite{Oshikawa-etal-2006}.  
In the phase diagram, the nontrivial phase occupies a lobe-shaped region at small \( {\Delta}_r \), occurring along the lines \( \phi = \pm 3\pi/2 \). Notably, the flux values where \( G_A \) reaches its maximum correspond precisely to the half-integer values \( \phi = (2n+1)\pi/2 \), for \( n \in \mathbb{Z} \), where Cooper-pair splitting dominates over paired Cooper-pair transport.

\begin{figure}[ht]
 	\centering
 	\includegraphics[width=0.47\linewidth]{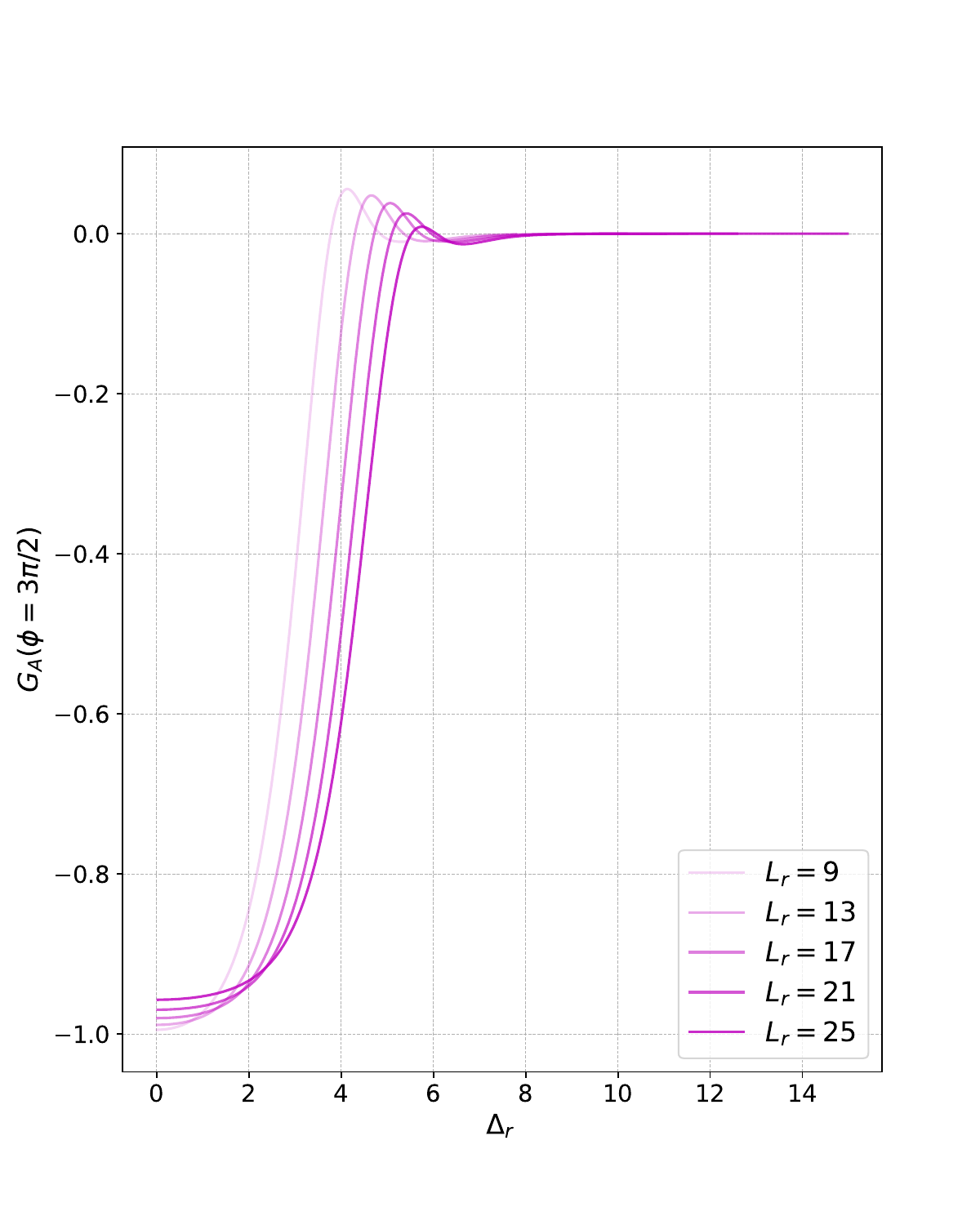}
 	\includegraphics[width=0.47\linewidth]{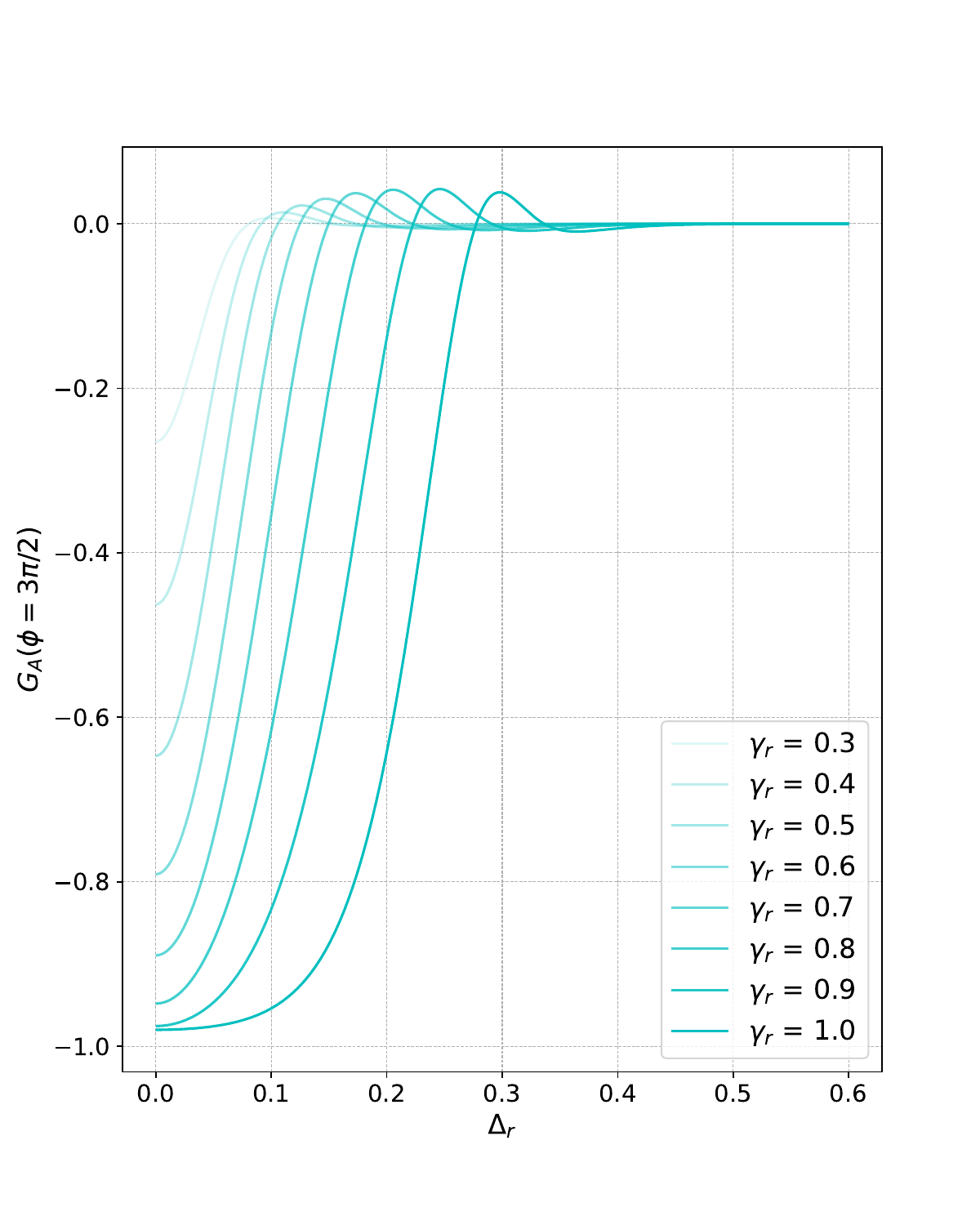}
 	\caption{Behavior of $G_A$ for fixed (left) system size $ L_r$ and (right) coupling $ \gamma_r$ 
  at fixed flux value $\phi = 3\pi/2$.}
 	\label{fig:gaat3pi2}
 \end{figure}

\begin{figure}[ht]
	\centering
	\includegraphics[width=0.47\linewidth]{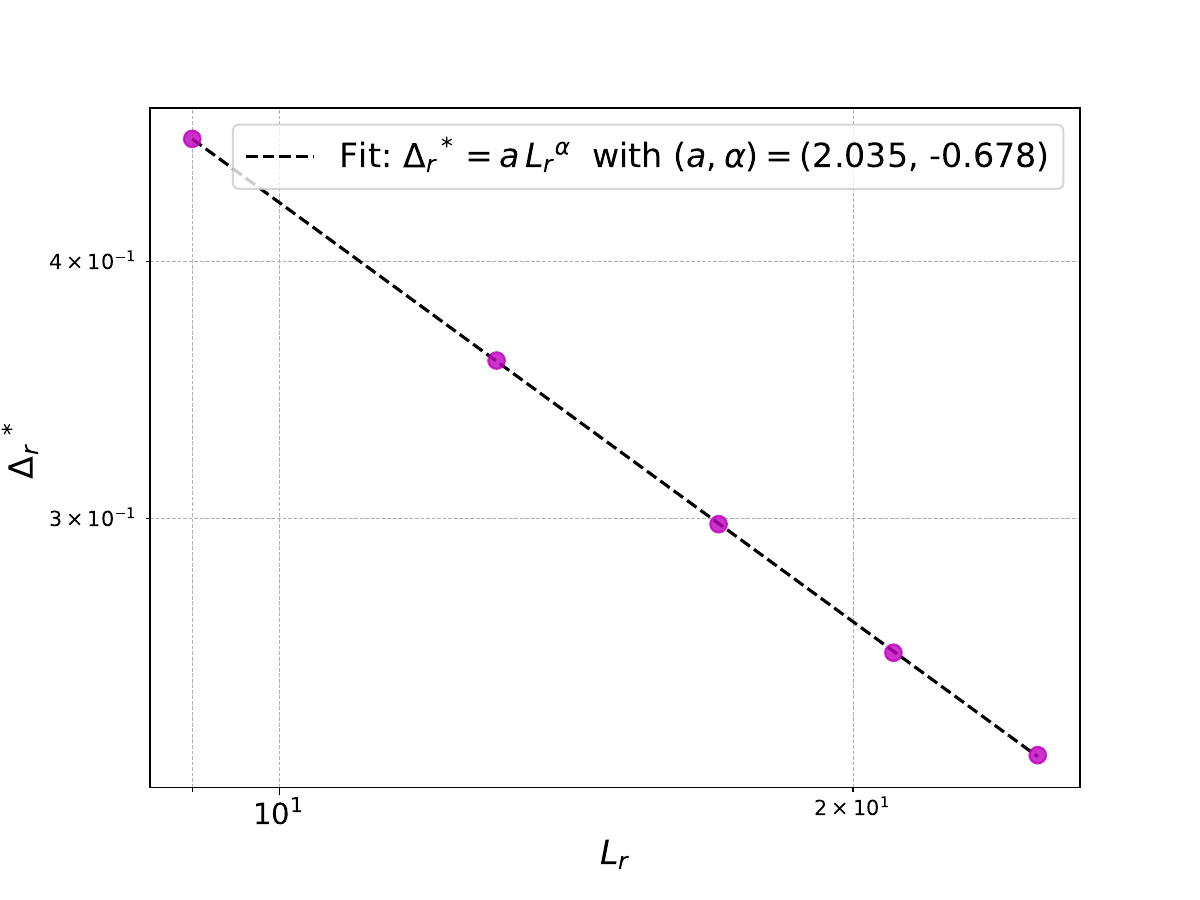}
	\includegraphics[width=0.47\linewidth]{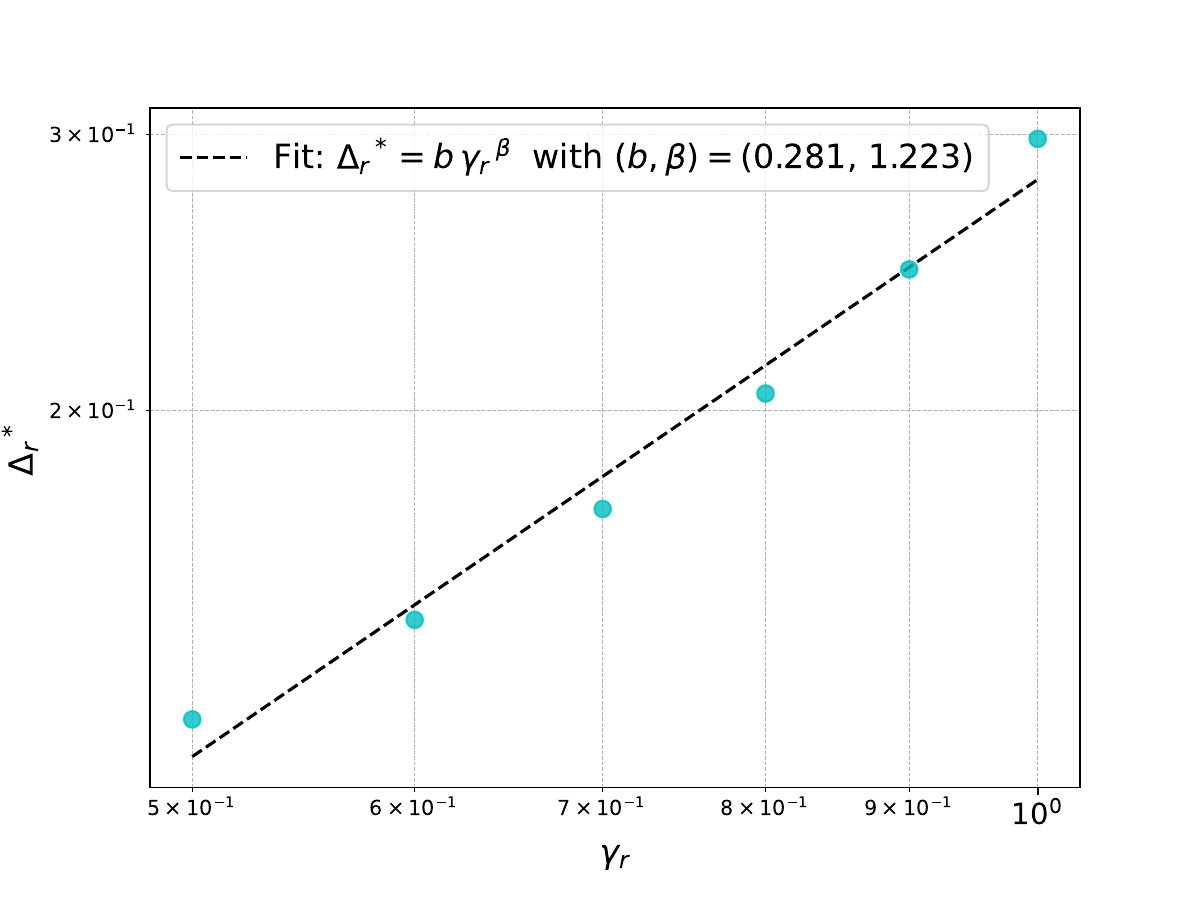}
	\caption{Peak locations of conductance in $\Delta_r$-space for varying parameters: (left) different system sizes
  and (right) different coupling strengths. In a log-log scale, the linear fit indicates a power law relation $ \Delta_r\sim  L_r^\alpha \gamma_r^\beta$.}
	\label{deltavslpdf}
\end{figure}

In the following, our goal is to extract the dependence of \( G_A \) on \( {\Delta}_r \), \( {\gamma}_r \), and \( {L}_r \), providing insight into the system’s behavior. Here, we focus on the line \( \phi = 3\pi/2 \), where Cooper-pair splitting is most prominent.   
The behavior of \( G_A \) for fixed \( {\gamma}_r \), with varying \( {L}_r \), and vice versa, is shown in Fig. \ref{fig:gaat3pi2}.  
We observe a decay from a large initial value of \( G_A \) (close to -1) toward zero, forming a peak at $ \Delta^*_r$ that marks the transition out of the Cooper-pair splitting regime. The behavior of \( G_A \) differs on either side of this point: for \( {\Delta}_r \gg {\Delta}^*_r \), it follows an exponential decay, whereas for smaller \( {\Delta}_r \), it exhibits an algebraic decay.  Fig. \ref{deltavslpdf} shows the location of the peak \( {\Delta}^*_r \) for different system sizes \( {L}_r \) and couplings \( {\gamma}_r \). In a log-log plot, the dependence of \( {\Delta}^*_r \) is well approximated by a power law. A linear fit to the data yields the relation  
\[
{\Delta}^*_r \sim {L}_r^\alpha {\gamma}_r^\beta,
\]  
with numerically determined exponents \( \alpha = -0.678 \) and \( \beta = 1.223 \).

\begin{figure}[ht]
	\centering
	\includegraphics[width=0.49\linewidth]{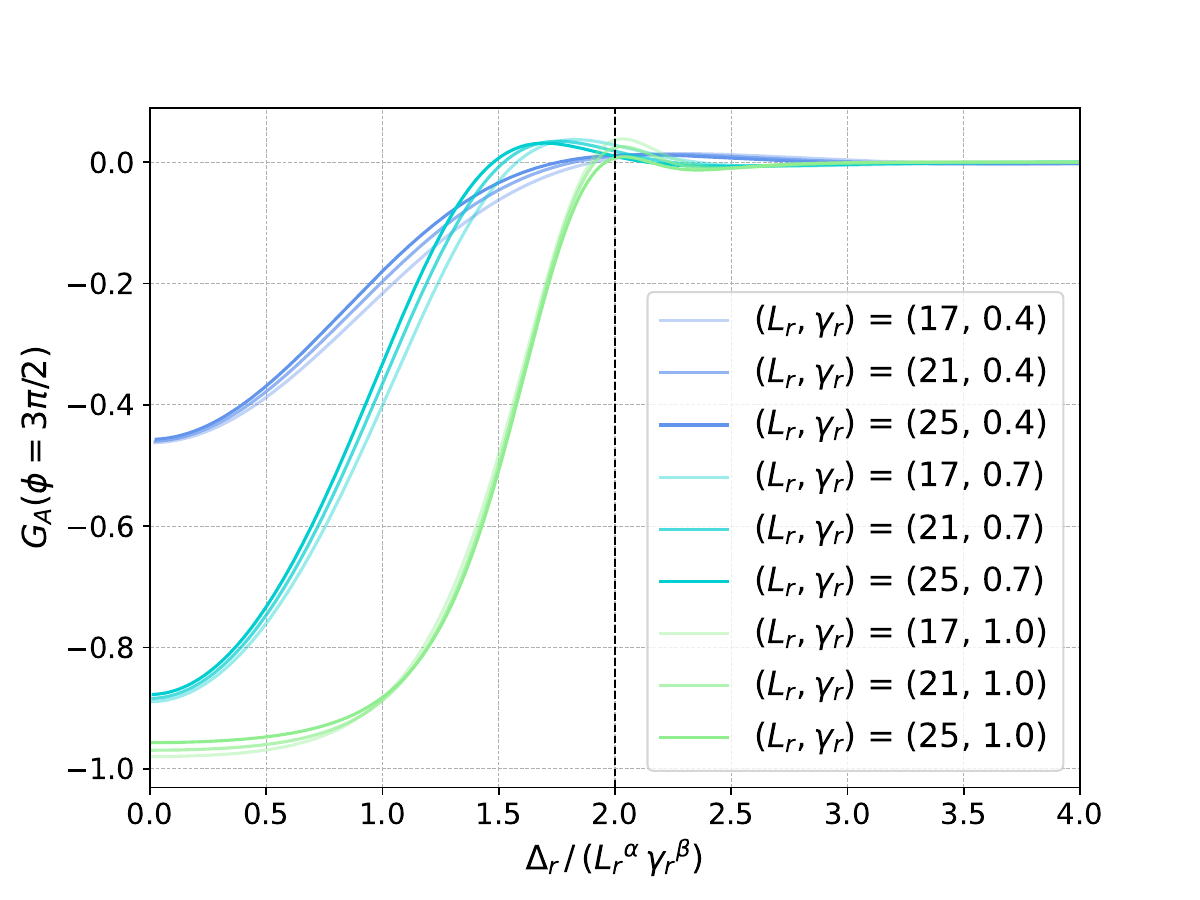}
		\includegraphics[width=0.49\linewidth]{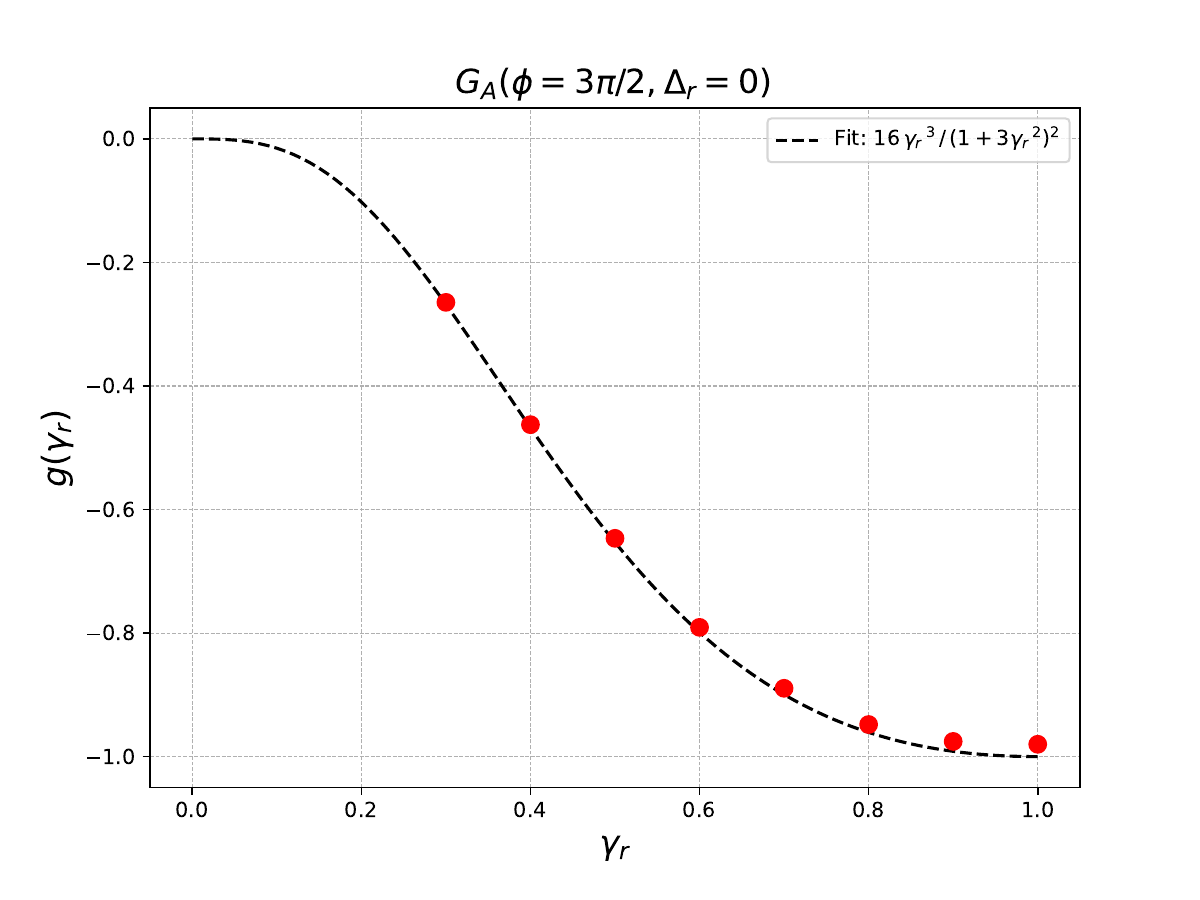}
 \caption{(Left) \( G_A \) for \( \phi = 3\pi/2 \) as a function of \( x \equiv {\Delta_r} / {L_r}^\alpha {\gamma_r}^\beta \) for various combinations of \( {L_r} \) and \( {\gamma_r} \). We observe (i) the collapse of the \( G_A \) peak around \( x \approx 2 \) and (ii) that \( G_A({\Delta_r} = 0) \) is largely insensitive to \( {L}_r \) and primarily depends on \( {\gamma_r} \). (Right) The behavior of \( G_A \) at \( {\Delta_r} = 0 \) is well fitted by the analytically derived function \( g({\gamma_r}) \).}
	\label{fig:gaat3pi2severallgamma}
\end{figure}

The left panel of Fig. \ref{fig:gaat3pi2severallgamma} shows the collapse of \( G_A \) in the early regime when plotted as a function of  
\[
x \equiv {\Delta_r} / ({L_r}^\alpha {\gamma_r}^\beta).
\]  
We observe that the peak occurs approximately at \( x \approx 2 \). Additionally, the data suggest that \( G_A \) is largely insensitive to the system size \( {L}_r \).  
This is further supported by the right panel, which displays the values of \( G_A \) for \( {\Delta}_r = 0 \), which are well fitted by the function  
\[ g( \gamma_r)\equiv \frac{G_A({\gamma_r}, \Delta_r=0,  L_r)}{e^2/h} = \frac{{\gamma_r}^3}{(1+3{\gamma_r}^2)^2}.
\]  
This expression was analytically derived for the case \( {L}_r = 1 \) in Appendix A of Ref. \cite{Oshikawa-etal-2006}.

\begin{figure}[ht]
	\centering
	\includegraphics[width=0.47\linewidth]{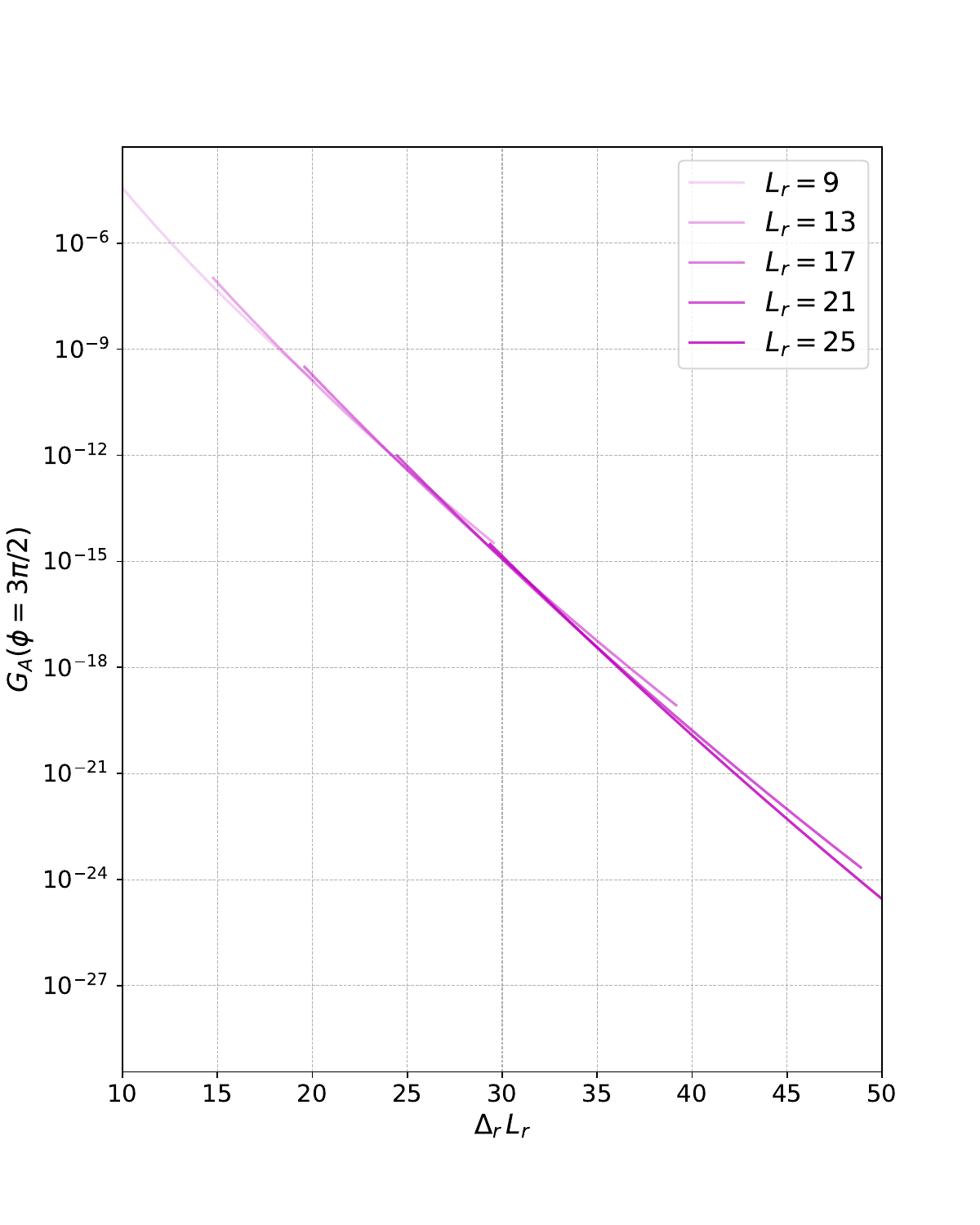}
	\includegraphics[width=0.47\linewidth]{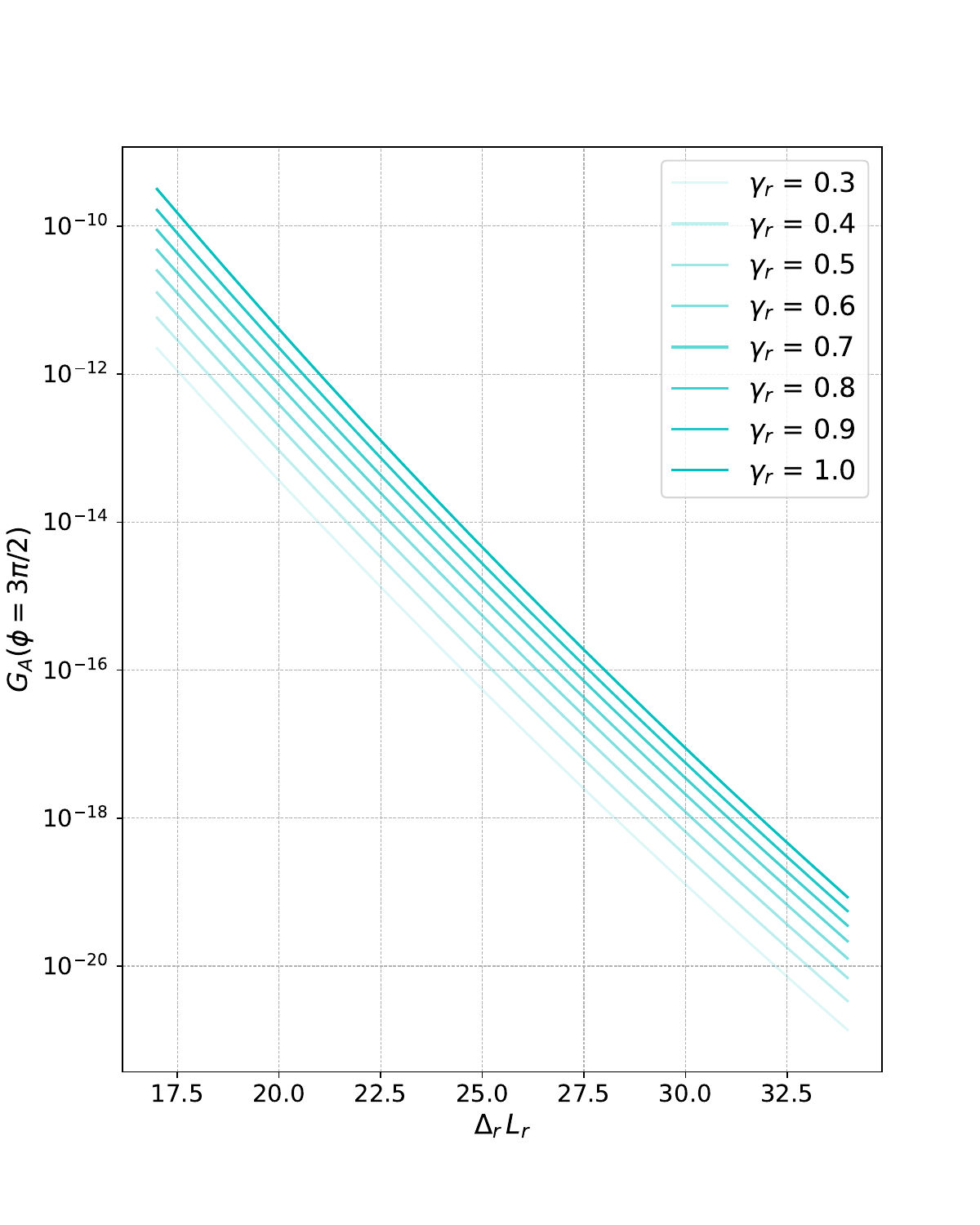}
	\caption{Exponential decay as a function of $\Delta_r\,  L_r $ for different (left) system sizes $ L_r$ and (right) hopping coupling $ \gamma_r$.}
	\label{fig:gaat3pi2vsgammaexptail}
\end{figure}

Finally, Fig. \ref{fig:gaat3pi2vsgammaexptail} reveals the exponential decay of \( G_A \) at large \( {\Delta_r} \). The exponential tail depends exclusively on the product \( {\Delta_r}\; {L_r} \) and is independent of \( {\gamma_r} \). This allows us to capture the general dependence of \( G_A \) along the line \( \phi = 3\pi/2 \) with the expression:  
\begin{eqnarray}
    \frac{G_A({\gamma_r}, {\Delta_r}, {L_r})}{e^2/h} = \frac{16\, {\gamma_r}^3}{(1+3{\gamma_r}^2)^2} f\left(\frac{{\Delta_r}}{{L_r}^\alpha {\gamma_r}^\beta}\right) \exp(-{\Delta_r} \, {L_r}),
\end{eqnarray}
where \( {\Delta_r} \), through the exponential term, introduces a correlation length \( \xi \equiv {a}\;{\Delta_r}^{-1} \) into the system.

The quantitative dependence of \( G_A \) on the system parameters is consistent with the qualitative description provided in Fig. 1 of the manuscript. In the limit \( {L_r} \gg {\Delta_r}^{-1} \), as shown in Fig. 1(b), the exponential decay dominates, causing \( G_A \) to approach zero -- reflecting the absence of pair splitting.  
Conversely, in the limit depicted in Fig. 1(c), where \( {\gamma_r}^\beta \ll {\Delta_r} \, {L_r}^\alpha \), the argument \( x \gg 2 \) of the function \( f(x) \) becomes too large, placing the system outside the relevant pair-splitting regime. This observation aligns with perturbative results, where pair splitting is a third-order process and is suppressed by the dominant second-order process in which electrons travel together.

\section{Tight-binding model for wire network}\label{molecule}

\begin{figure}[ht]
    \centering \includegraphics[scale=0.19]{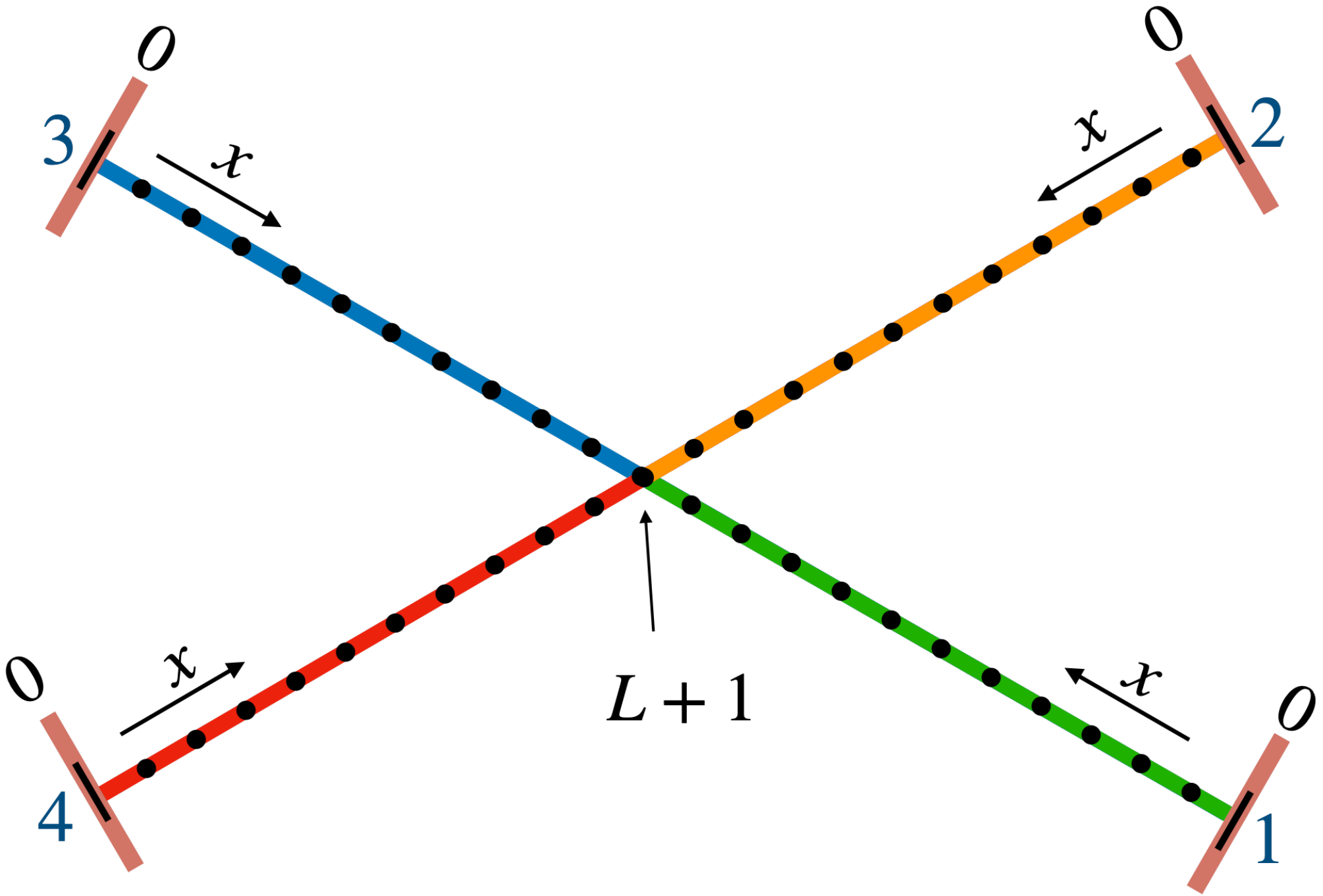}
    \caption{Position labels $x$ for sites along the four wire legs $1, 2, 3$, and $4$ (green, orange, blue, and red, respectively). Each coordinate lies in the range $x=0,\ldots, L+1$ and the four wires cross at $x =L+1$. }
    \label{labels}
\end{figure}

In this section we provide a detailed study of a model that captures
the finite length of the wires in the superconductor network. The
results in this section justify the effective model presented in
the main text, which treats each X-shaped section (``molecule") of the network as a single electron degree of freedom. Here, we consider a tight-binding description of the
network, where each wire segment is composed of multiple sites, each
containing an electron degree of freedom.The barriers, that become Josephson junctions in the strong superconducting pairing limit, are
modeled by weak links through which electrons tunnel.
The unit cell, shown in Fig. \ref{fig:unitcell}, consists of three
copies of X-molecules (displayed individually in
Fig.~\ref{labels}) coupled to each other through a barrier,
indicated in red. The unit cell thus contains $3(4L+1)$ atoms, and we
note that the effective model studied in the main body of the paper
corresponds to the extreme limit of $L=0$.

Let $c^\dagger_{\textbf{r},a,i,\sigma}$ be the spin $\sigma = \uparrow, \downarrow$ electron creation
operator at (kagome lattice) site $\mathbf{r}$ and sublattice $a=1,2,3$, where $i=1,\ldots,4L+1$ labels the sites inside each X-molecule (Fig.~\ref{labels}). The electron
operators obey the anti-commutation relation
$\{c^\dagger_{\textbf{r},a,i,\sigma},c^{}_{\textbf{r}',b,j,\sigma'}\}
= \delta_{\mathbf{r}, \mathbf{r}'}\, \delta_{a,b}\, \delta_{i,j}\,
\delta_{\sigma\sigma'}$. The Bogoliubov-des Gennes (BdG) quadratic
Hamiltonian consists of four terms
\begin{align}\label{N_fermions}
  H = H_t + H_\gamma + H_\mu+H_\Delta
  \;,
\end{align}
corresponding, respectively, to intra-X hoping of amplitude $t$,
inter-X hopping with amplitude
$\gamma\le t$, chemical potential $\mu$, and an on-site  superconducting
pairing with amplitude $\Delta$. Explicitly, these terms are given by
\begin{align}
  H_t &=-t \sum_{\mathbf{r}}\sum_{a=1,2,3} \;\sum_{\sigma=\uparrow,\downarrow}
        \;\sum_{\langle i,j\rangle\in \text{X}}\left(\; c^{\dagger}_{\mathbf{r}, a, i,\sigma}\,c^{\,}_{\mathbf{r},a, j,\sigma} + \text{h.c.}\right)
        \;,\\
  H_\gamma&=\gamma \sum_{\mathbf{r}} \sum_{\sigma=\uparrow, \downarrow} \sum_{\langle i, j\rangle \in \partial \mathbf{X}}\left[\sum_{a=1,2,3} e^{i \phi / 3}\left(c_{\mathbf{r}, a, i, \sigma}^{\dagger} c_{\mathbf{r}, a+1, j, \sigma}+c_{\mathbf{r}, a, i, \sigma}^{\dagger} c_{\mathbf{r}+\mathbf{v}_a, a+1, j, \sigma}\right)+\text { h.c. }\right]
    \;,\\
  H_\mu &= -\mu \sum_{\mathbf{r}} \sum_{a=1,2,3} \; \sum_{\sigma=\uparrow,\downarrow}
          \;\sum_{i \in \text{X}} \left (c^{\dagger}_{\mathbf{r}, a, i,\sigma}\,c^{}_{\mathbf{r}, a, i,\sigma} + \text{h.c.}\right)
          \;,\\
  H_{\Delta} &= \sum_{\mathbf{r}}\sum_{a=1,2,3} 
               \;\sum_{i\in \text{X}} \left(\Delta^{}_{a, i}\; c^{\dagger}_{\mathbf{r}, a, i,\uparrow}\,c^{\dagger}_{\mathbf{r},a, i,\downarrow} + \text{h.c.}\right)
               \;,
\end{align}
where $\mathbf v_{a}= \mathbf s_a - \mathbf s_{a+1}$.
The notation in the above $\langle i,j\rangle\in \text{X}$ refers to a
sum over nearest neighbors inside the X-molecule (as in
Fig. \ref{labels}), while $\langle i,j\rangle\in\partial \text{X}$ refers to
a sum over nearest neighbors at the tips of distinct wires, through the junction (red links) in Fig. \ref{fig:unitcell}. Additionally, the label $a$ in $c_{\mathbf{r}, a, i, \sigma}$ is only defined mod $3$. The
Hamiltonian in Eq. \eqref{N_fermions}, expressed in momentum space,
is a $2\times 3\times (4L+1)$ dimensional matrix. In the following we study
this momentum space Hamiltonian and argue that, in certain regimes, its low-energy physics is completely captured by the effective minimal model studied in the main body of the paper.

\begin{figure}[ht]
    \centering \includegraphics[scale=0.25]{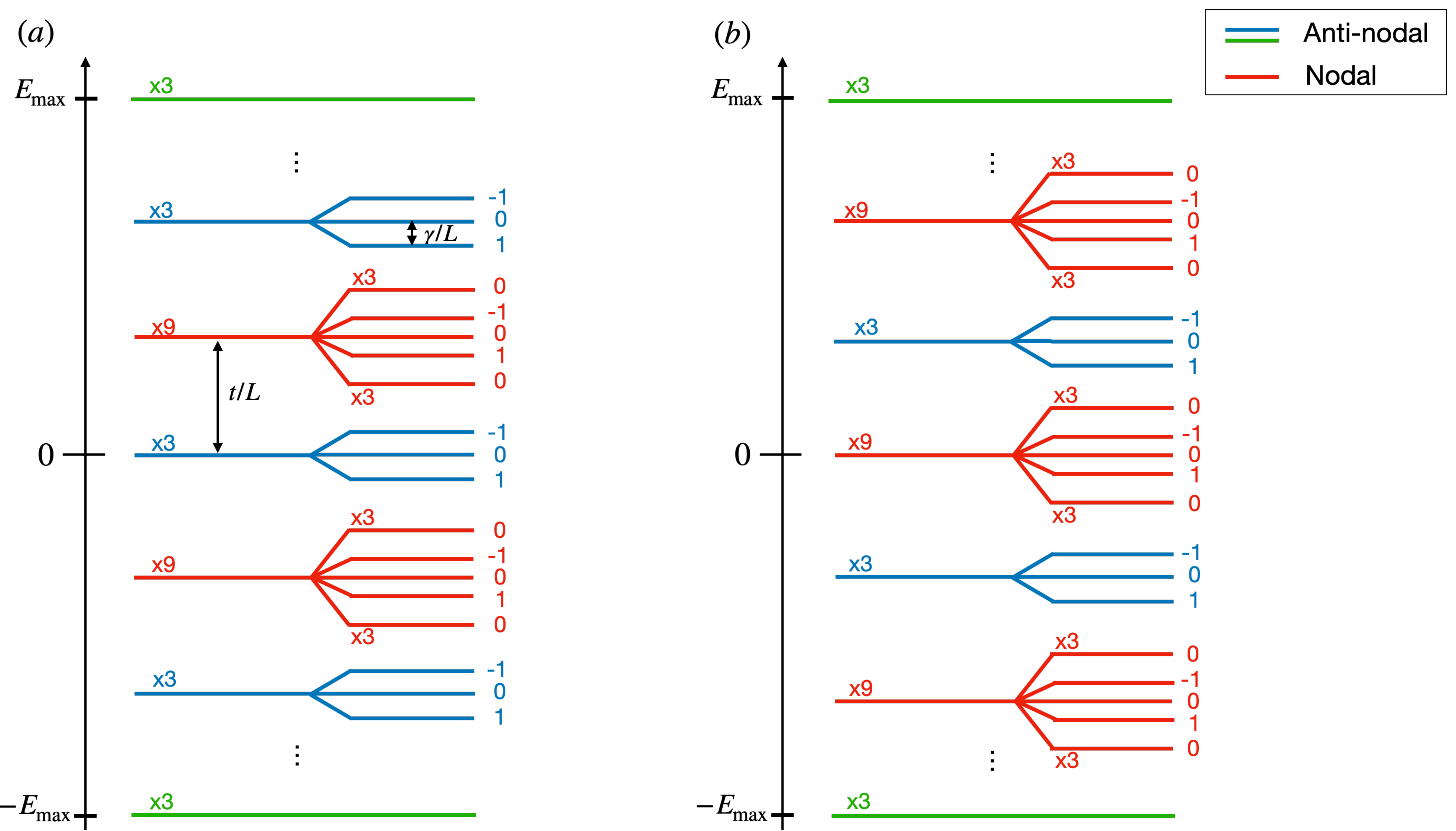}
    \caption{Spectrum of the Hamiltonian in Eq.~\eqref{N_fermions} for $\Delta=0$ and $\mu=0$. Pattern of nodal and anti-nodal flat bands, depicted in red and blue respectively for $\gamma=0$ when (a) $L$ is even and (b) $L$ is odd. The states at extreme ends of the spectrum have energy $\pm E_{\text{max}}$, are depicted in green, and correspond to anti-nodal states with imaginary momentum Eq.~\eqref{k_imaginary}. The degeneracy for each band (per spin) is indicated with an x. The figure also show the splitting when inter X-molecule hopping $\gamma\neq 0$, with Chern numbers of each band indicated on the right for $\phi=\pi/2$. The bands with imaginary momentum (green) have an exponentially small splitting (in $L$) in the presence of $\gamma$, and the individual bands also present non-trivial Chern numbers $1, 0, -1$, similar to other anti-nodal clusters (the splitting in the figure is not shown for clarity). }
    \label{fig:bands_pattern}
\end{figure}

It is useful to start in the limit $\gamma=0$, where the
X-molecules decouple. In this limit the spectrum has no
momentum dependence and all bands are flat, with each band labeled by
the eigenstates of the isolated molecule orbitals. For each spin
state, the spectrum is shown in Fig.~\ref{fig:bands_pattern} and consists
of alternating clusters of 3-fold and
$(3\times 3)$-fold degenerate states and two additional states at the extremes. Which cluster sits at zero energy is determined by whether $L$ is even or odd. The $(4L+1)$ states in the spectrum
(matching the number of sites in the unit cell) are accounted for as
follows.

\begin{figure}[ht]
    \centering \includegraphics[scale=0.35]{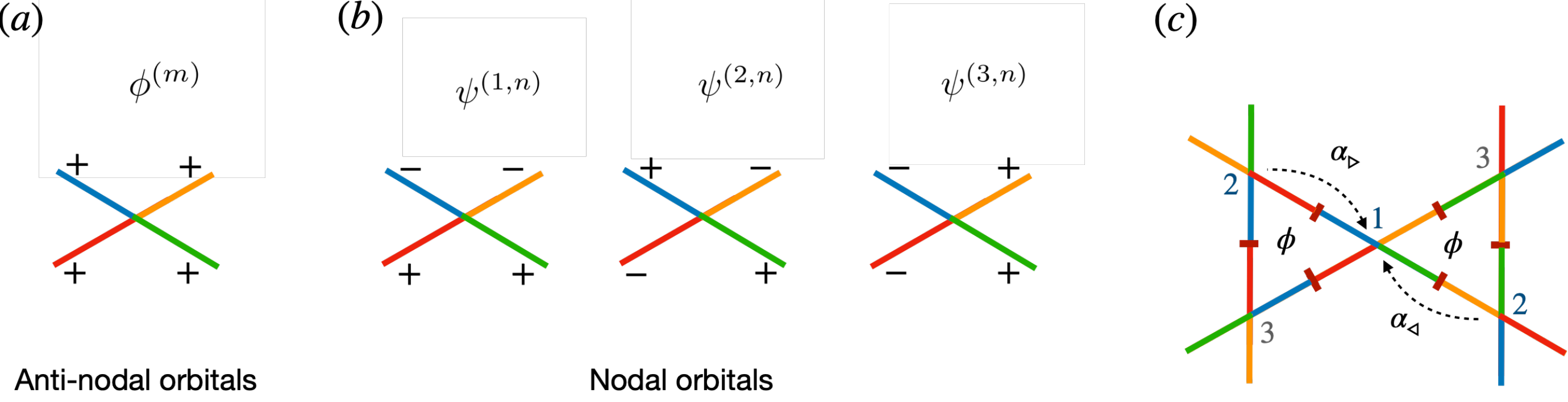}
    \caption{Wave-function configurations for (a) anti-nodal and (b) nodal orbitals. $\pm$ denotes the relative sign, up to a gauge choice, of the wave function on each leg. The colors green, orange, blue, and red label, respectively, the  $1,2,3,$ and $4$ legs of the X-molecules. (c) The matrix elements $(\alpha_{\triangleright})_{rs}$ and $(\alpha_{\triangleleft})_{rs}$  are constructed from the hopping between $\psi^{(r,n)}$ and $\psi^{(s,n)}$ orbitals along the three sub-lattice sites $1,2$, and $3$ in the kagome lattice. The crossings are labeled by the same numbers as the sub-lattice sites 1, 2, and 3.}
    \label{orbitals}
\end{figure}

{\it States with a node at the center of the X-molecule}: These states
correspond to eigenvectors of the X-molecule tight binding model which vanish at the crossing point. We label the sites by a discrete variable $x$ that runs inward starting from $x=0$ at
the edge of each leg and ending at the crossing point $x=L+1$, as shown in Fig.~\ref{labels}. The following is a solution of the Schr\"odinger equation in real space: 
\begin{eqnarray}
    \psi^{(n)}(x)=+{\mathcal A}_{L}\,\sin\left(\frac{\pi}{L+1}n\,x\right)
\end{eqnarray}
with normalization factor $\mathcal A_{L} = 1/{\sqrt{2L+2}}$, and $n=1,\ldots,L$ in each of the four legs.
Boundary conditions require that $\psi^{(n)}(0) =\psi^{(n)}(L+1)=0$. There are three such nodal solutions $\psi^{(r,n)}_j$ labeled by $r=1,2,3$ on each leg $j=1,2,3,4$ which are of the form
\begin{align}
  &\psi^{(1,n)}_1(x) = -\psi^{(1,n)}_2(x)=-\psi^{(1,n)}_3(x)= \psi^{(1,n)}_4(x)=\psi^{(n)}(x),
  \nonumber\\
  &\psi^{(2,n)}_1(x) = -\psi^{(2,n)}_2(x)=\psi^{(2,n)}_3(x)= -\psi^{(2,n)}_4(x)=\psi^{(n)}(x),
  \nonumber\\
  &\psi^{(3,n)}_1(x) = \psi^{(3,n)}_2(x)=-\psi^{(3,n)}_3(x)= -\psi^{(3,n)}_4(x)=\psi^{(n)}(x).
  \label{psi_nodal}
\end{align}
The shape of the wavefunction is identical in each leg except that two legs have opposite sign in each of the three cases. Intuitively, each of these solutions corresponds to the usual particle hopping in one-dimension glued together on two intersecting lines such that the amplitude is zero at the crossing. The orbitals are orthogonal and are illustrated in
Fig.~\ref{orbitals} (b). The eigenvalues are
\begin{align}
  E_n = -2t\,\cos\left(\frac{\pi}{L+1}n\right),
  \quad
  n=1,\dots, L
  \;.
\end{align}
We refer to these solutions as \textit{nodal} and they correspond to $3L$ states in our $(4L+1)$-dimensional Hilbert space per X-molecule. For $L$ odd the solution with $n=(L+1)/2$ is at zero energy.

{\it States with non-zero amplitude at the center of the X-molecule}: There are additional solutions that do not vanish at the crossing point and are symmetric with respect to all the four legs,
 which account for the remaining $L+1$ states. Consider the ansatz
\begin{align}
  \phi^{(m)}(x)={\mathcal B}_{m,L}\,\sin\left(q_m\,x\right)\,,
\end{align}
where $\mathcal B_{m,L}$ depends on both $L$ and a quantum number $m$, and  $q_m$ is a parameter to be determined. We will enumerate $m$ below.  Consider further the ansatz on each of the four legs:
\begin{eqnarray}\label{anti-nodal}
    \phi_1^{(m)}(x) = \phi_2^{(m)}(x) = \phi_3^{(m)}(x) = \phi_4^{(m)}(x) = \phi^{(m)}(x).
\end{eqnarray}
This form satisfies the equation of motion at all sites other than the crossing. The energy of these modes is
\begin{align}
  E_m = -2t\,\cos\left(q_m\right).
  \;
  \label{E_antinodal}
\end{align}
There is an additional boundary condition at the crossing which constrains the values of $q_m$:
\begin{align}\label{energies_anti_nodal}
  -t\left[\phi^{(m)}_1(L)+\phi^{(m)}_2(L)+\phi^{(m)}_3(L)+\phi^{(m)}_4(L)\right]
  =
  E_m \;\phi^{(m)}(L+1).
\end{align}
Using the ansatz in Eq.~\eqref{anti-nodal} and the condition in Eq.~\eqref{energies_anti_nodal} implies that
\begin{eqnarray}
  \label{junction}
  2\sin(q_mL)= \cos(q_m)\;\sin\left(q_m(L+1)\right)
  \;.
\end{eqnarray}
We find $L-1$ solutions for $q_m$ with $m=1,\ldots,L-1$ where $q_m$ is real. Additionally, there are two imaginary solutions
\begin{eqnarray}\label{k_imaginary}
q_0=\mathrm i \kappa\quad \text{and}\quad q_L = \pi+\mathrm i \kappa
\end{eqnarray}
with energies $\pm  E_{\text{max}} = \pm2t\cosh(\kappa)$ and $\kappa$ real. (These energies correspond to the lowest and largest eigenvalues shown in Fig.~\ref{fig:bands_pattern}.) The wavefunctions for these states decay exponentially in magnitude as one moves away from the crossing. For $L$ even there is a zero energy solution, labeled by $m_0$, such that $q_{m_0}=\pi/2$ satisfies both Eq.~\eqref{E_antinodal} and \eqref{junction}.  To summarize, we have accounted for all $4L+1$ modes by solving explicitly for $3L$ nodal and $L+1$ anti-nodal solutions.

As the inter-molecule tunneling $\gamma$ is turned on, the degeneracy in the nodal and anti-nodal orbitals split, forming a band structure of alternating clusters. The energy splitting due to the presence of $\gamma$ is illustrated in Fig. \ref{fig:bands_pattern} (a) and (b) at $(k_x,k_y)=(0,0)$. We show the band structure along high-symmetry points in Fig. \ref{fig:moleculebands}. Even at $\gamma\neq 0$ some of the bands remain flat and gapped for semi-integer flux values (namely the middle bands in the anti-nodal and nodal clusters and the outer bands in the nodal clusters). Interestingly, we have verified numerically that even for $\gamma\sim t$, there are no band crossings, i.e., the anti-nodal and nodal clusters do not mix together and are shielded by the outer bands in nodal clusters.

We now argue that the physics at small $\gamma$ for $\mu$ near zero is dominated by the three bands around zero energy regardless the parity of $L$, which justifies the effective Hamiltonian in Eq.~\eqref{eq:effective-H}. First, consider the case where $L$ is even, where the states located at zero energy are anti-nodal. The effective Hamiltonian $H^{\operatorname{eff}, \operatorname{A}}$ is a $3\times 3$ matrix as in Eq.~\eqref{H_Kagome} where the matrix elements are determined by the hopping of the zero energy orbital state $\phi^{(m_0)}$ in Eq.~\eqref{anti-nodal} between X-molecules:
\begin{align}
H_{\mathbf{k}}^{\operatorname{eff}, \operatorname{A}} = \gamma
\begin{bmatrix}
    0 & \alpha^{\,}_{\triangleright} + \alpha^{\,}_{\triangleleft}\,d_{\mathbf{k}}^{12} & \bar\alpha^{\phantom\dagger}_{\triangleright} + \bar\alpha^{\phantom\dagger}_{\triangleleft}\,d_{\mathbf{k}}^{13} \\
    \bar\alpha^{\phantom\dagger}_{\triangleright} + \bar\alpha^{\phantom\dagger}_{\triangleleft}\,d_{\mathbf{k}}^{21} & 0 & \alpha^{\,}_{\triangleright} + \alpha^{\,}_{\triangleleft}\,d_{\mathbf{k}}^{23} \\
    \alpha^{\,}_{\triangleright} + \alpha^{\,}_{\triangleleft}\,d_{\mathbf{k}}^{31} & \bar\alpha^{\phantom\dagger}_{\triangleright} + \bar\alpha^{\phantom\dagger}_{\triangleleft}\,d_{\mathbf{k}}^{32} & 0
\end{bmatrix}
\;,
\end{align}
where
\begin{eqnarray}
 \alpha^{\,}_{\triangleright} = e^{i\phi/3} \phi^{(m_0)}_2( 1)\phi^{(m_0)}_1(1) =\mathcal B_{m_0,L}^2 \, e^{i\phi/3} \quad \text{and} \quad
 \alpha^{\,}_{\triangleleft} = e^{i\phi/3} \phi^{(m_0)}_3( 1)\phi^{(m_0)}_4(1) = \mathcal B_{m_0,L}^2\, e^{i\phi/3}\;.
\end{eqnarray}
Upon turning on superconductivity $\Delta$, we recover the effective model in Eq.~\eqref{eq:effective-H}.

Second, consider $L$ odd. At $\gamma=0$ there are nine degenerate nodal states at zero energy in the unit cell. The effective Hamiltonian in this degenerate subspace results in a Hamiltonian $H^{\operatorname{eff}, \operatorname{N}}$ which is a $9\times 9$ matrix:
\begin{align}
H_{\mathbf{k}}^{\operatorname{eff}, \operatorname{N}} = \gamma
\begin{bmatrix}
    0_{3\times 3} & \alpha^{\,}_{\triangleright} + \alpha^{\,}_{\triangleleft}\,d_{\mathbf{k}}^{12} & \alpha^{\dagger}_{\triangleright} + \alpha^{\dagger}_{\triangleleft}\,d_{\mathbf{k}}^{13} \\
    \alpha^{\dagger}_{\triangleright} + \alpha^{\dagger}_{\triangleleft}\,d_{\mathbf{k}}^{21} & 0_{3\times 3} & \alpha^{\,}_{\triangleright} + \alpha^{\,}_{\triangleleft}\,d_{\mathbf{k}}^{23} \\
    \alpha^{\,}_{\triangleright} + \alpha^{\,}_{\triangleleft}\,d_{\mathbf{k}}^{31} & \alpha^{\dagger}_{\triangleright} + \alpha^{\dagger}_{\triangleleft}\,d_{\mathbf{k}}^{32} & 0_{3\times 3}
\end{bmatrix}
\;,
\label{nodal_overlap}
\end{align}
where $\alpha_{\triangleright}$ and $\alpha_{\triangleleft}$ are now $3\times 3$ matrices that are determined by hopping between the nodal orbitals in Eqn. \eqref{psi_nodal} on adjacent X-molecules. Explicitly, 
\begin{eqnarray}
    \alpha_{\triangleright} =\mathcal A_{L}^2\, e^{i\phi/3} \left( \begin{matrix}
-1 & +1 & +1\\
+1 & -1 & -1\\
-1 & +1 & +1
    \end{matrix}\right)\quad \text{and}\quad \alpha_{\triangleleft} = \mathcal A_{L}^2\, e^{i\phi/3} \left( \begin{matrix}
-1 & -1 & +1\\
-1 & -1 & +1\\
-1 & -1 & +1
    \end{matrix}\right).\label{alpha_matrix}
\end{eqnarray}
As an example of how to construct the $\alpha_{\triangleright}$ and $\alpha_{\triangleleft}$ matrices, consider the hopping between the zero energy orbitals $\psi^{(1,n_0)}$ and $\psi^{(2,n_0)}$ orbitals, as defined in Eq. \eqref{psi_nodal} with $n_0=(L+1)/2$,  which accounts for the matrix elements $(\alpha_{\triangleright})_{12}$ and $(\alpha_{\triangleleft})_{12}$. As shown in Fig. \ref{orbitals}(c), $(\alpha_{\triangleright})_{12}$ is the result of hopping between leg 3 (blue) of molecule 1 and leg 4 (red) of molecule 2. From Fig. \ref{orbitals}(b) we see that orbital $\psi^{(1,n_0)}_3$ has amplitude $-\mathcal A_{L}$ and $\psi^{(2,n_0)}_4$ has amplitude $-\mathcal A_{L}$, so their product is
\begin{eqnarray}
(\alpha_{\triangleright})_{12}=e^{i\phi/3} \, \psi^{(1,n_0)}_3(1)\, \psi^{(2,n_0)}_4(1)=\mathcal A_{L}^2\, e^{i \phi/3}.
\end{eqnarray}
Similarly, $\alpha_{\triangleleft}$ is the result of hopping between leg 1 (green) of molecule 1 and leg 2 (orange) of molecule 2, and it is given by  
\begin{eqnarray}
(\alpha_{\triangleleft})_{12}=e^{i\phi/3} \, \psi^{(1,n_0)}_1(1)\, \psi^{(2,n_0)}_2(1)= -\mathcal A_{L}^2\, e^{i \phi/3}.
\end{eqnarray}
The remaining matrix elements for $\alpha_{\triangleright}$ and $ \alpha_\triangleleft$ are constructed following the same procedure.

\begin{figure}[ht]
    \centering
    \includegraphics[scale=0.39]{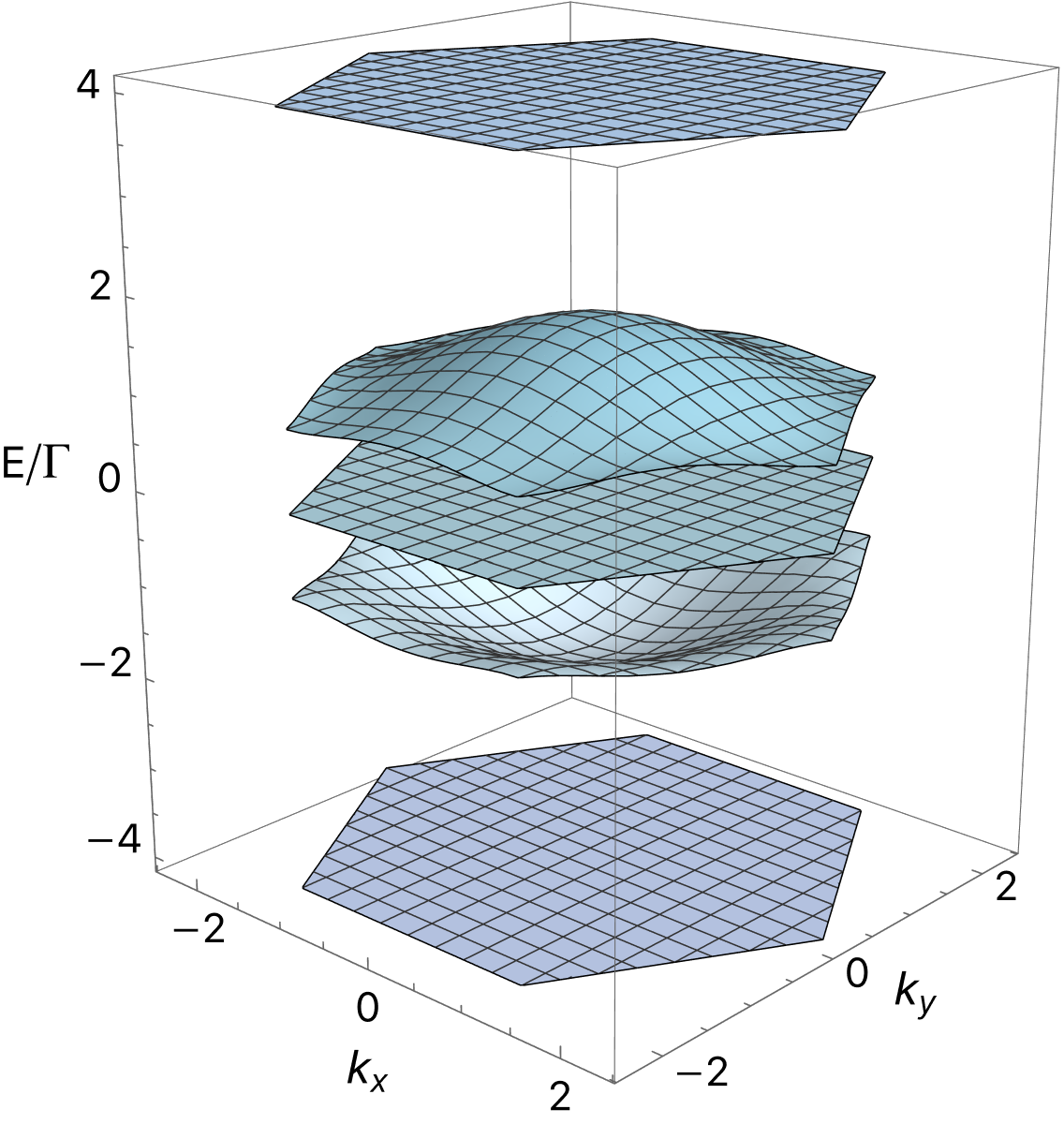}
    \caption{Band structure of $H_{\mathbf{k}}^{\operatorname{eff}, \operatorname{N}}$, with $\Gamma = \gamma\, \mathcal A_{L}^2$. The flat bands at energies $\pm 4 \Gamma$ are 3-fold degenerate. The three bands in the middle correspond to the effective model in Eq. \eqref{eq:effective-H}.}
    \label{fig:nodal_bands}
\end{figure}

Upon diagonalization of $H_{\mathbf{k}}^{\operatorname{eff}, \operatorname{N}}$ we observe that a small $\gamma$ splits the 9 bands into two sets of three-fold degenerate flat bands with zero Chern number at energies $\pm 4 \Gamma$, where $\Gamma =\gamma \mathcal A_L^2$, and three other bands sandwiched between them and separated by a gap (Fig.~\ref{fig:nodal_bands}). We also observe that the eigenvectors in the middle bands consist only of superpositions of the orbitals $\psi^{(2,n_0)}_j$ in Eq.~\eqref{psi_nodal}, where opposite legs have the same sign. Keeping only these orbitals in the Hamiltonian $H_{\mathbf{k}}^{\operatorname{eff}, \operatorname{N}}$ gives us:
\begin{eqnarray}
    \alpha_{\triangleright} = \alpha_{\triangleleft} = \, \mathcal A_{L}^2\, e^{i\phi/3} \left( \begin{matrix}
0 & 0 & 0\\
0 & 1 & 0\\
0 & 0 & 0
    \end{matrix}\right).
\end{eqnarray}

The three bands in the middle carry Chern numbers $+1$, $0$ and $-1$.
Thus we conclude that these three bands reproduce the effective Hamiltonian per spin $h_{\mathbf{k}}$ in the effective theory Eq.~\eqref{eq:effective-H}. 
Note that since the 3-fold degenerate flat bands at $\pm 4\Gamma$ have vanishing Chern numbers, they do not contribute to the Hall conductivity at half-filling.

Our previous derivation indicates that regardless of the parity of $L$, ensured one is at half filling $\mu=0$, the low-energy physics is always determined by the effective Hamiltonian in Eq.~\eqref{eq:effective-H}. Not surprisingly, we find non-trivial topological numbers for appropriate values of flux. Since the physics in the middle bands are determined by the effective model, Chern numbers assume non-vanishing values for $\phi$ around half-integer multiples of $\pi$, $\phi_p=(p-1/2)\pi$ for $p\in \mathbb Z$, with opposite sign depending whether $p$ is either odd or even. The pattern of Chern numbers for all the bands is displayed in Fig. \ref{fig:bands_pattern} (a) and (b) for $\phi=\pi/2$, where we note that the degenerate flat bands in the nodal clusters have vanishing Chern-number. As one moves away from such special flux values $\phi_p$, by an amount $\sim \pm\pi/4$, the bands acquire enough curvature to start crossing with each other. The electrons at the Fermi sea then acquire dispersion and present a metallic behavior, signaling a phase transition to the non-topological regime. 
Finally upon turning on superconductivity $\Delta\neq 0$, if we ensure that the chemical potential obeys $-|\Delta|<\mu<|\Delta|$, then a topological superconducting phase is achieved.

\subsection*{Relevant length and energy scales to reach chiral topological superconductivity}

The regime in which chiral topological superconductivity emerges in the effective model (see Fig.~\ref{fig:orderpar}) corresponds to large ratios of the hopping $\Gamma$  to the superconducting amplitude $\Delta$. Here we use the tight-binding model described above to relate this ratio $\Gamma/\Delta$ to the ratio between the coherence length and the size of the triangles in the kagome lattice.

The Fermi velocity within the X-molecule is $v_F = 2 t a$, where $a$ is the lattice spacing and $t$ the hopping amplitude. The physical (dimensionfull) length $\ell$ of the wires (within the X-molecule) can be expressed as $\ell = (2L+1) a$. The spacing between the eigenmodes of the X-molecules is order $v_F/\ell \sim t/L$ (as illustrated in Fig.~\ref{fig:bands_pattern}).

The hopping amplitude for the effective model with one site per X-molecule can be seen from the above discussion to be $\Gamma = \gamma 
\mathcal B_{m_0,L}^2$ or $\gamma \mathcal A_{L}^2$ depending on whether $L$ is even or odd, respectively. Both the amplitudes $\mathcal B_{m_0,L}$ and $\mathcal A_{L}$ scale with the inverse of  the square root of ${L}$. Thus, for both $L$ even and odd $\Gamma \sim \gamma/L$. In the regime in which the tunnel barriers between the X-molecules are small, i.e., when they are almost transparent so as to favor the case shown in Fig.~\ref{fig:Y-splitters}(a), we can use $\gamma\sim t$, so that $\Gamma \sim t/L \sim v_F/\ell$. Finally, using the definition of the superconducting coherence length $\xi=v_F/\Delta$, we can express
\begin{align}
    \frac{\Gamma}{\Delta}\sim \frac{\xi}{\ell}
    \;,
\end{align}
and therefore the regime in which we encounter chiral topological superconductivity in Fig.~\ref{fig:orderpar} corresponds to the case in which the coherence length is larger than the size of the triangles in the kagome lattice, as intuitively expected.

\section{Josephson junction limit}\label{classical}

In the limit $\Delta\gg\Gamma$ the low-energy physics is dominated by the Cooper-pair condensate and pair splitting effects are suppressed. In such a regime, we expect the effective Hamiltonian to be a function of only the superconducting order parameter and the magnetic flux. 
We assume translation invariance such that the superconducting order parameter repeats in each unit cell, $\Delta_{\mathbf{r},a} = \Delta e^{i\theta_a}$. Looking at Fig.~\ref{orbitals}(c), every hopping between sublattice sites $a$ and $b$ contains two factors: a Josephson tunneling energy proportional to $\cos(\theta_a-\theta_b)$ and a Peierls factor $e^{2i\phi}$ for the clockwise tunneling of each Copper pair. This gives us:
 \begin{eqnarray}
	H_{\text{JJ}} = -E_J\left [\cos\left (\theta_2 - \theta_1{+}\dfrac{2\phi}{3}\right )+\cos\left (\theta_1 - \theta_3 {+} \dfrac{2\phi}{3}\right )+\cos\left (\theta_3 - \theta_2 {+} \dfrac{2\phi}{3}\right )\right ],\label{JJ}
\end{eqnarray}
 where $E_J=\Gamma^2/|\Delta|$ is the Josephson energy scale of the array. The ground state of $H_{\text{JJ}}$ is protected by a superconducting gap of order $ \Delta$, and its energy is minimized by different configurations of $\theta_a$ depending on the magnetic flux $\phi$.

The following three phase configurations carry different chirality $\chi$ (see Eq.~\eqref{eq:chirality}) and are noteworthy:
\begin{eqnarray}
	\chi = 0\quad &\Rightarrow&\quad(\theta_1, \theta_2, \theta_3) = (0,0,0) \nonumber\\
	\chi=1\quad &\Rightarrow&\quad (\theta_1, \theta_2, \theta_3) = \left (0, {2\pi}/{3}, -{2\pi}/{3}\right )\nonumber\\
	\chi=-1\quad &\Rightarrow&\quad (\theta_1, \theta_2, \theta_3) = \left (0, -{2\pi}/{3}, {2\pi}/{3}\right ).\label{chiral_config}
\end{eqnarray}
As we will see, these are the minima of $H_{\text{JJ}}$ depending on the value of $\phi$.
Since only phase differences are physical, we can fix $\theta_1=0$. 

\begin{figure}[ht]
	\centering
	\includegraphics[width=0.5\linewidth]{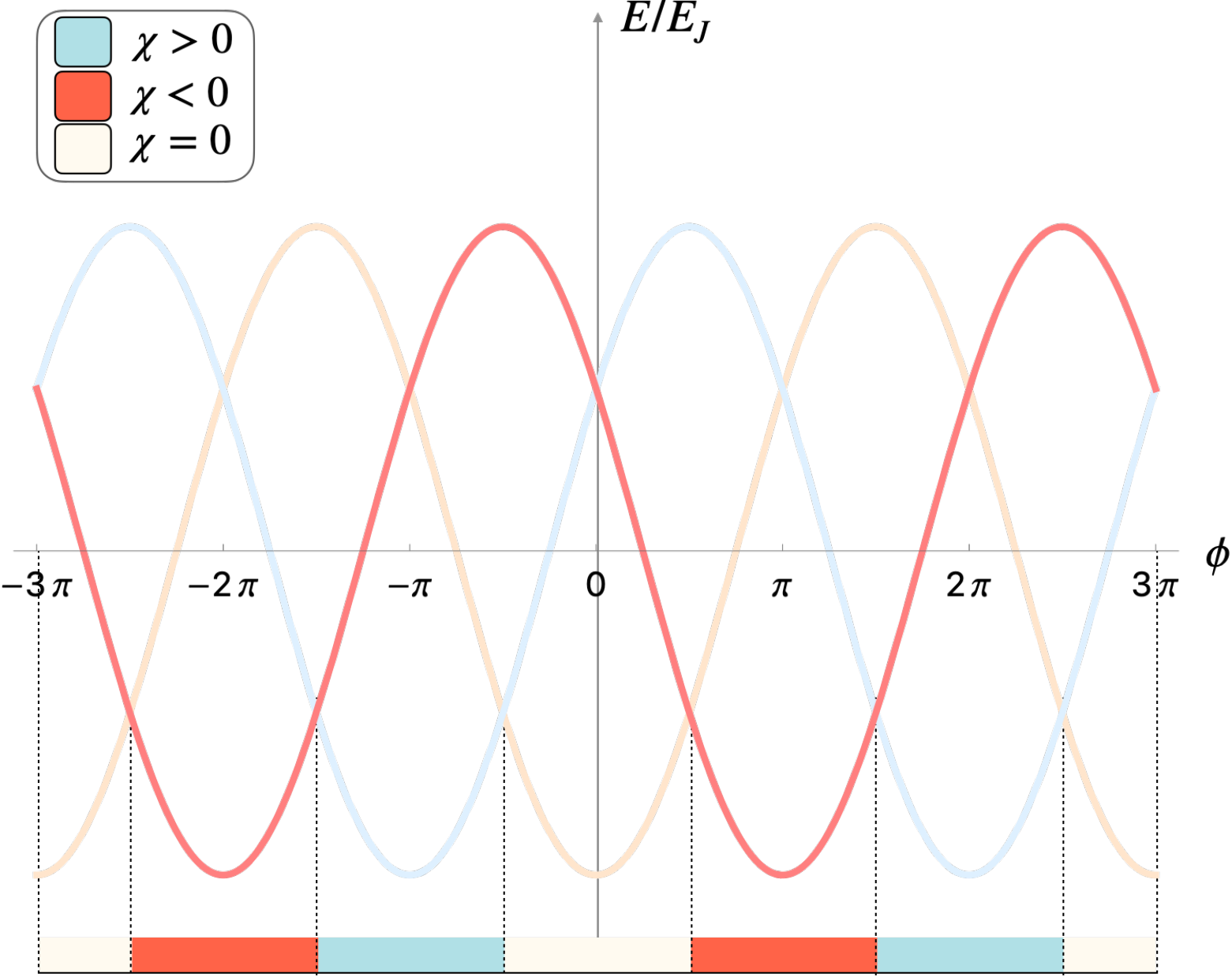}
	\caption{Mean field energy associated with different $\theta_a$ configurations. The chirality $\chi$ of $\theta_a$ configurations that minimize the energy for different values of flux $\phi$ agree with the limit of small $\Gamma/\Delta$ in Fig. \ref{fig:orderpar} (colored pattern with different $\chi$ at the bottom of the figure).}
	\label{fig:largedeltalimit}
\end{figure}

In Fig. \ref{fig:largedeltalimit} we show the Josephson energy in Eq.  \eqref{JJ} for the three different phase configurations. There are three intercalating regions where one configuration in Eq. \eqref{chiral_config} at a  time minimizes the energy. This behavior agrees with the numerical quantum results that were presented in the lower part of the phase diagram in Fig. \ref{fig:orderpar} for $\Gamma/\Delta< 1$. The periodicity of the phase diagram in Fig. \ref{fig:largedeltalimit} is $6\pi$, in contrast to the usual $2\pi$. This can be seen directly from the classical expression in Eq.~\eqref{JJ} and is a result of Cooper pair formation. One can explicitly check that for $\Delta=0$, the Hamiltonian in Eq.~\eqref{eq:effective-H} is $2\pi$ periodic.

\bibliography{references}


\end{document}